\documentclass[reprint,superscriptaddress,preprintnumbers,nofootinbib,amssymb,aps]{revtex4-1}

\usepackage[utf8]{inputenc}
\usepackage[colorlinks=true,citecolor=blue,linkcolor=blue]{hyperref}
\usepackage[normalem]{ulem}
\usepackage{amsmath,amssymb,mathtools}
\usepackage{epsfig}
\usepackage{graphicx}               
\usepackage{url}
\usepackage{color,colortbl}
\usepackage{slashed}
\usepackage{cancel}
\usepackage{multirow}
\usepackage{placeins}
\usepackage[dvipsnames]{xcolor}
\usepackage{epstopdf}
\usepackage{soul}
\usepackage[capitalise, english]{cleveref}
\usepackage{siunitx}
\usepackage{xspace}
\usepackage{makecell}

\definecolor{red}{rgb}{0.6,.0706,.1373}
\definecolor{blue}{rgb}{0,0.396,0.741}
\newcommand\myshade{80}
\colorlet{mylinkcolor}{violet}
\colorlet{mycitecolor}{violet}
\colorlet{myurlcolor}{violet}

\hypersetup{
  linkcolor  = mylinkcolor!\myshade!black,
  citecolor  = mycitecolor!\myshade!black,
  urlcolor   = myurlcolor!\myshade!black,
  colorlinks = true
}

\allowdisplaybreaks

\newcommand{\be}{\begin{equation}}
\newcommand{\ee}{\end{equation}}
\newcommand{\bea}{\begin{eqnarray}}
\newcommand{\eea}{\end{eqnarray}}

\def\beq#1\eeq{\begin{align}#1\end{align}}

\def\thesection{\Roman{section}}

\makeatletter
\providecommand*{\diff}%
  {\@ifnextchar^{\DIfF}{\DIfF^{}}}
\def\DIfF^#1{%
  \mathop{\mathrm{\mathstrut d}}%
    \nolimits^{#1}\gobblespace}
\def\gobblespace{%
  \futurelet\diffarg\opspace}
\def\opspace{%
  \let\DiffSpace\!%
  \ifx\diffarg(%
    \let\DiffSpace\relax
  \else
    \ifx\diffarg[%
      \let\DiffSpace\relax
    \else
        \ifx\diffarg\{%
        \let\DiffSpace\relax
      \fi\fi\fi\DiffSpace}


\begin{document}

\title{
The Dark Side of a Tera-Z Factory
}

\author{Pablo Olgoso}
\email{pablo.olgosoruiz@unipd.it}
\affiliation{Dipartimento di Fisica e Astronomia ”Galileo Galilei”, Universit`a di Padova, Italy}
\affiliation{Istituto Nazionale di Fisica Nucleare (INFN), Sezione di Padova, Via F. Marzolo 8, 35131 Padova, Italy}
\author{Paride Paradisi}
\email{paride.paradisi@pd.infn.it}
\affiliation{Dipartimento di Fisica e Astronomia ”Galileo Galilei”, Universit`a di Padova, Italy}
\affiliation{Istituto Nazionale di Fisica Nucleare (INFN), Sezione di Padova, Via F. Marzolo 8, 35131 Padova, Italy}
\author{Nud\v{z}eim Selimovi\'c\,}
\email{nudzeim.selimovic@pd.infn.it}
\affiliation{Istituto Nazionale di Fisica Nucleare (INFN), Sezione di Padova, Via F. Marzolo 8, 35131 Padova, Italy}


\preprint{}

\begin{abstract}
The future circular $e^+e^-$ collider (FCC-ee or CEPC) will provide unprecedented sensitivity to indirect new physics signals emerging as small deviations from the Standard Model predictions in electroweak precision tests. Assuming new physics scenarios containing a dark matter candidate and a $t$-channel mediator, we analyse the synergy and interplay of future Tera-$Z$ factories and non-collider tests conducted through direct and indirect searches of dark matter. Our results highlight the excellent prospect for a Tera-$Z$ run to indirectly probe the presence and nature of dark matter. 
\end{abstract}

\maketitle

\section{Introduction} 
\label{sec:intro}

The nature and origin of dark matter (DM) remain among the most pressing open questions in physics. So far, cosmological and astrophysical observations have firmly established only general requirements that any viable DM candidate has to satisfy. In particular, DM should be stable on cosmological scales, electrically neutral, non-relativistic at the epoch of matter-radiation equality, and with self-interactions compatible with the observations of cluster collisions, such as the Bullet Cluster.
Moreover, any DM model has to comply with the increasingly stringent bounds from direct detection (DD) as well as indirect detection (ID) searches of DM.

In the popular WIMP scenario, it is assumed that the DM candidates were in thermal equilibrium in the very
early stages of the Universe and, at later times,
decoupled (froze-out) from the primordial plasma. 
In this framework, and assuming a standard cosmological history for the Universe, the DM relic density is fully determined by the thermally averaged pair annihilation cross-section. One of the most appealing features of this scenario relies on the fact that the DM relic density requires sizable couplings between the Standard Model (SM) and DM particles, making such DM models potentially testable at particle accelerators and low-energy physics experiments.

Therefore, to make progress in uncovering the DM particle properties, it is essential to explore how the dark sector may couple to the visible, SM sector~\cite{Cirelli:2024ssz}. 
A broad class of such interactions is captured by so-called portals, including the Higgs portal~\cite{Patt:2006fw,OConnell:2006rsp,Krnjaic:2015mbs,Arcadi:2019lka,Haisch:2023aiz,Ghosh:2025dcv}, kinetic mixing with a dark photon~\cite{Holdom:1985ag,Fayet:1990wx,Dobrescu:2004wz,delAguila:1995rb,Graham:2021ggy}, axion-like particle portals~\cite{Nomura:2008ru,Gola:2021abm,Bharucha:2022lty,Ghosh:2023tyz,Dror:2023fyd,Armando:2023zwz,Allen:2024ndv,DEramo:2025xef}, portals with heavy neutral lepton~\cite{Blennow:2019fhy,Gorbunov:2007ak,Alekhin:2015byh}, or topological portals~\cite{Davighi:2024zip,Davighi:2025awm}, among others. 

These portals are often studied by means of simplified models, \emph{i.e.}, minimal extensions of the SM in terms of the number of new particles and couplings introduced.
One of the simplest realizations is represented by the so-called $s$-channel portals, where the SM is extended by the DM candidate and a spin-$0,1$ mediator coupled to DM and SM fermion pairs. The absence of interactions with an odd number of DM particles is typically ensured by imposing that the DM (but not the mediator) is charged under a discrete $\mathbb{Z}_2$ symmetry. 
Moreover, both the DM and the mediator are assumed to be singlets under the SM gauge groups. Simplified models with $s$-channel mediators exhibit a very strong complementarity between the DM relic density and the expected signals at DD and ID searches~\cite{Arcadi:2024ukq}.

In this work, we focus on scenarios in which \emph{all} dark sector particles are charged under a $\mathbb{Z}_2$ symmetry to ensure DM stability. Then, SM–DM interactions naturally arise through $t$-channel exchange of the mediator.
Contrary to the case of $s$-channel simplified models, the thermally averaged cross-section and therefore also the DM relic density are now very sensitive to coannihilation processes associated with the $t$-channel mediator.

In this respect, we first classify all possible interactions at the renormalisable level in which the DM particle is a singlet under the SM gauge group. Even if simplified, their rich phenomenology has been studied in the literature~\cite{Garny:2015wea,giacchino2016signatures,baek2016top,Garny:2018icg,colucci2018top,Arina:2020tuw,Arina:2023msd,Bertuzzo:2024bwy,Arina:2025zpi,Roy:2025pht,Angel:2025xwb,Reis:2025avx,Belanger:2025wjh,Biondini:2025gpg}, and they could describe a large class of more complete dark sector constructions. 

Due to the $\mathbb{Z}_2$ symmetry, such $t$-channel portals can affect SM processes only via one-loop corrections. Therefore, these contributions are typically small and hard to isolate in current experimental data. Recently, however, it has been recognized that potential next-generation high-luminosity facilities like FCC~\cite{FCC:2025lpp}, CEPC~\cite{CEPCStudyGroup:2018ghi}, LEP3~\cite{Anastopoulos:2025jyh}, 
Muon Collider~\cite{Accettura:2023ked}, ILC~\cite{ILC:2013jhg}, CLIC~\cite{CLICdp:2018cto}, LCF~\cite{LinearCollider:2025lya}, will possess the precision required to probe new physics (NP) quantum effects, with dedicated studies exploring this possibility in various settings~\cite{deBlas:2021jlt,deBlas:2022ofj,Kamenik:2023hvi,Celada:2024mcf,Allwicher:2024sso,Erdelyi:2024sls,Greljo:2024ytg,Gargalionis:2024jaw,Maura:2024zxz,Davighi:2024syj,Erdelyi:2025axy,terHoeve:2025gey,Maura:2025rcv,Allwicher:2025bub,Azzi:2025dwl,Greljo:2025ggc,Cornet-Gomez:2025jot}.

Among future facilities, Tera-$Z$ factories such as FCC-ee stand out in their ability to constrain loop-induced effects with unprecedented precision. Thus, this article aims to assess the potential of the Tera-$Z$ program to probe $t$-channel DM portals, highlighting the role of these searches in constraining broad classes of $\mathbb{Z}_2$-symmetric models. In our analysis, we emphasize the complementarity between collider experiments, underground facilities, and cosmological probes, such as the interplay between the Tera-$Z$ runs, the DARWIN~\cite{DARWIN:2016hyl} direct detection experiment, and the Cherenkov Telescope Array Observatory (CTAO)~\cite{CTAConsortium:2013ofs}. 

While next-generation non-collider experiments such as DARWIN and CTAO are expected to begin operations before the Tera-$Z$ collider, it is worth emphasising that a deeper understanding of (possible) underlying new physics effects would highly benefit from the complementary information achievable at a Tera-$Z$ factory. This is due to the unique ability of a Tera-$Z$ to probe correlations between precision observables that are not accessible
in direct or indirect DM searches alone. 

The physics program outlined in this work is expected to represent a relevant and timely contribution in the context of ongoing discussions on future directions in particle physics.

\section{\texorpdfstring{$\mathbf{t}-$Channel Models}{t-Channel Models}}
\label{sec:setup}
For DM candidates charged under the SM gauge group, the annihilation cross section is typically determined by electroweak interactions, which dominate over channels mediated by heavier states. Moreover, direct detection experiments strongly constrain DM candidates with non-zero hypercharge ($Y \neq 0$). A well-known scenario involving a $\mathrm{SU}(2)_L$ multiplet with $Y = 0$ is Minimal Dark Matter~\cite{Cirelli:2005uq,Cirelli:2007xd,Cirelli:2009uv,Hambye:2009pw,Cohen:2011ec,Kumericki:2012bh}, with a DM mass around 10 TeV. Such heavy candidates remain largely unconstrained at the FCC-ee, motivating our focus on scenarios in which DM is a complete SM singlet.

Under the assumptions of renormalizable interactions and no more than two new fields, this framework excludes $s$-channel portal models, with the exception of the Higgs portal for scalar DM. However, $t-$channel mediator models remain viable and present interesting phenomenology at future facilities. In these scenarios, the $\mathbb{Z}_2$ symmetry that stabilizes the DM ensures that dark sector effects are absent at tree level, leading only to loop-induced corrections in the visible sector.

Depending on whether the DM candidate is a fermion ($\chi$) or a scalar ($\phi$), and on the spin of the $t-$channel mediator (a fermion $\Psi$ or a scalar $\Phi$), various coupling structures to SM particles arise. We categorize these scenarios using the schematic notation “DM–Mediator–SM”. Since DM is a total singlet under the SM gauge group, the $t-$channel mediator must transform under the same representation as the SM particle it couples to. All feasible portals are summarized in Tab.~\ref{table:models}. The representative scenarios in each category (entry in Tab.~\ref{table:models}) are discussed in Sec.~\ref{sec:results}.

Among all these portals, there are two --namely $\chi\Psi H$ and $\phi\Phi H$-- which feature a significant difference: after electroweak symmetry breaking, the DM candidate mixes with the neutral component of the heavy mediator. This mass mixing induces, in turn, an interaction of DM to the $Z$ boson proportional to the coupling to the Higgs.
This mediates a tree-level scattering with nucleons, which forces such coupling to be $\sim\mathcal{O}(0.1)$ in order to avoid exclusion by current direct detection experiments. Thus, these scenarios remain unconstrained at a Tera-$Z$ factory and will be omitted in the following.

So far, we have made no reference to the nature of the fermion DM candidate. If $\chi$ is a Dirac fermion, the interaction with the $t$-channel mediator generates an effective coupling to the $Z$-boson at one-loop which, again, is subject to stringent constraints from direct detection experiments. 
Evading these limits typically renders the coupling too weak to be probed at a Tera-$Z$ factory. In contrast, if $\chi$ is a Majorana fermion, such interaction is absent, which motivates us to pursue this scenario in the following.

The interaction Lagrangian for the portals with a Majorana DM candidate $\chi$ can be collectively written as
\begin{equation}
    -\mathcal{L}_{\chi}=\kappa |\Phi|^2 |H|^2 +\left(y_\psi \overline{\chi}\psi \Phi^\dagger +{\rm h.c.}\right)\,,
    \label{eq:Lag_MajoranaDM}
\end{equation}
where $\psi$ stands for any SM fermion and $y_\psi$ is a vector in flavor space. In addition, portals with $\mathrm{SU}(2)_L$-doublet mediators also include the following coupling
\begin{equation}
    -\mathcal{L}_{\chi}\supset \kappa_2 (\Phi^\dagger \sigma^I \Phi) (H^\dagger \sigma^I H)\,,
     \label{eq:kappa_2}
\end{equation}
where $\sigma^I$ are the Pauli matrices. In the case of a scalar singlet DM candidate $\phi$, the Lagrangian can be written as
\begin{equation}
    -\mathcal{L}_{\phi}= \kappa |\phi|^2 |H|^2 + \left(y_\psi \overline{\Psi}\psi \phi + {\rm h.c.}\right)\,.
     \label{eq:Lag_ScalarDM}
\end{equation}

\begin{table}[t]
\renewcommand{\arraystretch}{2} 
\setlength{\tabcolsep}{12pt} 
\centering
\begin{tabular}{|c|c|c|}
\cline{1-3}
\cline{1-3}
Mediator & DM spin $\frac{1}{2}$: $\chi$ & DM spin $0$: $\phi$ \\
\cline{1-3}
spin $\frac{1}{2}$: $\Psi$ & $\chi\,\Psi\,H$ & 
$\phi\,\Psi\,\psi$
\\
\hline
spin $0$: $\Phi$ & 
$\chi\,\Phi\,\psi$
& $\phi\, \Phi\, H$ 
\\
\hline
\end{tabular}
\caption{All possible $t$-channel portals involving a scalar ($\phi$) or fermionic ($\chi$) dark matter candidate and a scalar ($\Phi$) or fermionic ($\Psi$) mediator. The SM Higgs is denoted by $H$, while the set of SM fermions is collectively represented as $\psi = \{q_L, u_R, d_R, \ell_L, e_R\}$.}
\label{table:models}
\end{table}

\section{Phenomenological Setup}
\label{sec:pheno}

Before analyzing the representative scenarios within each class of viable $t$-channel mediator models, we first outline the phenomenological framework used in our study. 
Our analysis primarily relies on precision observables measured at the $Z$-pole during the anticipated Tera-$Z$ run of future lepton colliders, complemented by potential signatures in both direct and indirect dark matter detection experiments.

\subsection{\texorpdfstring{Tera-$Z$}{Tera-Z}} 
\label{sec:teraZ}

The main sensitivity of future $e^+e^-$ colliders to $t$-channel dark matter portals originates from the $Z$-pole run, owing to the production of around $10^{12}$ $Z$ bosons. To quantify the expected constraints, we construct a $\chi^2$ function of the form
\begin{equation}
\chi_{\rm TeraZ}^2 = \sum_{ij}[O_{i,{\rm exp}} - O_{i,{\rm th}}] (\sigma^{-2})_{ij}[O_{j,{\rm exp}} - O_{j,{\rm th}}]\,,
\label{eq:chi2_EWPO}
\end{equation}
where $O_i$ denote the measurable observables, and $\sigma^{-2}$ is the inverse of the associated covariance matrix.

We focus on the electroweak sector in the $\{\alpha_{\rm EM}, m_Z, G_F\}$ input scheme, considering the following electroweak precision observables (EWPOs): $\{\Gamma_Z$, $\sigma_{\rm had}$, $R_f$, $A_f$, $A_{\rm FB}^{0,\ell}$, $A_c^{\rm FB}$, $R_{uc}$, $m_W$, $\Gamma_W$, {\rm Br}$(W\to \ell \nu)$, $R_{W_c}$, $A_b^{\rm FB}\}$. Experimental inputs and SM predictions are taken from Refs.~\cite{ALEPH:2005ab, ALEPH:2013dgf, Janot:2019oyi, dEnterria:2020cgt, SLD:2000jop, ParticleDataGroup:2020ssz, CDF:2005bdv, LHCb:2016zpq, ATLAS:2016nqi, D0:1999bqi, ATLAS:2020xea}, following the methodology of Refs.~\cite{Breso-Pla:2021qoe,Allwicher:2023aql}. The projected experimental uncertainties for a Tera-$Z$ factory are obtained by rescaling current measurements using the state-of-the-art projections published in the latest Feasibility Study Report for FCC-ee~\cite{FCC:2025lpp,Altmann:2025feg,Belloni:2022due}. Throughout this analysis, we assume that future measurements are centered on the SM predictions.

In addition to the electroweak precision observables, we incorporate information from the $240$ GeV run, optimized for studying the properties of the Higgs boson produced in association with a $Z$ boson. Thus, we account for potential new physics effects in the Higgstrahlung production cross-section and in the Higgs signal strengths. However, the impact of Higgs observables becomes significant only at large values of the Higgs portal couplings $\kappa$ and $\kappa_2$ in Eqs.~\eqref{eq:Lag_MajoranaDM} and~\eqref{eq:kappa_2}, respectively. As a result, their influence on the overall picture is limited across most of the parameter space, where the constraints remain dominated by the $Z$-pole observables.

To identify which electroweak precision observables primarily drive the constraints at future Tera-$Z$ facilities in the portals under consideration, we introduce the pull of a given observable $O$ as
\begin{equation}
P_O = \frac{\delta O_{\rm NP}}{\sigma_{O,{\rm exp}}}\,,
\label{eq:pulls}
\end{equation}
where $\delta O_{\rm NP}$ represents the shift in the observable induced by new physics, such that the theoretical prediction takes the form $O_{\rm th} = O_{\rm SM} + \delta O_{\rm NP}$. Here, $O_{\rm SM}$ denotes the SM prediction, and $\sigma_{O,{\rm exp}}$ is the expected experimental precision for $O$ at the Tera-$Z$ stage of future lepton colliders. For each portal, we identify two observables with the largest pull and examine their dependence on the relevant model parameters. Importantly, the characteristic pattern of deviations observed in different $t$-channel portals can serve as a way to discriminate between them.

A particularly important aspect of our results is that the leading constraints on $t$-channel portals often stem from the branching ratio of the $Z$ boson into bottom quarks $R_b$, and the $b$-quark forward-backward asymmetry $A_{\rm FB}^b$. This is largely due to their anticipated experimental precision gains, as detailed in the latest FCC-ee Feasibility Study Report~\cite{FCC:2025lpp}, and their dominant role in driving new physics pulls in the global electroweak fit. In scenarios involving portal to quarks, $R_b$ typically provides the strongest sensitivity, while $A_{\rm FB}^b$, being sensitive to the electron–$Z$ coupling, becomes relevant also in leptonic cases. Achieving the projected level of sensitivity, however, relies on non-trivial progress in SM calculations. Throughout this study, we adopt an optimistic assumption where theoretical uncertainties do not spoil the experimental ones.

In this context, we note the approach adopted in studies such as~\cite{Freitas:2019bre,EWPPGTH}, which assess the impact of theoretical uncertainties on electroweak precision observables by defining scenarios of theory improvement. In a similar way, our work identifies the specific observables where improved SM calculations would yield the greatest gains in sensitivity for probing dark matter at the Tera-$Z$. This highlights concrete priorities for future theoretical efforts to ensure that the full potential of the experimental program can be realised.

Lastly, we assume that both the DM candidate and the $t$-channel mediator are heavy and, as such, we can describe the relevant physics at low energies using the Standard Model Effective Field Theory (SMEFT)~\cite{Brivio:2017vri}. To that end, one needs to integrate out the heavy particles, with the leading effects appearing at one loop due to $\mathbb{Z}_2$ symmetry and quadratic couplings of the heavy particles. For each model, we use \texttt{SOLD}~\cite{Guedes:2023azv,Guedes:2024vuf,Fonseca:2020vke} and \texttt{matchmakereft}~\cite{Carmona:2021xtq} to obtain the complete one-loop matching onto the SMEFT.

\subsection{Relic Abundance}
\label{sec:relic_abundance}

Our goal is to assess how a Tera-$Z$ program can probe the parameter space of $t$-channel dark matter portals relevant for the observed relic abundance. We focus on the standard scenario in which a $\mathbb{Z}_2$-odd dark sector remains in chemical equilibrium with the SM in the early universe, and the DM abundance is set by thermal freeze-out.

In the models of interest, DM annihilates via the $t$-channel process $\mathrm{DM}\,\mathrm{DM} \to \overline{\psi}\psi$, where $\psi$ denotes the corresponding SM fermion (see Tab.~\ref{table:models}). The thermally averaged annihilation cross section $\langle \sigma v \rangle$ is helicity-suppressed, making next-to-leading-order (NLO) processes particularly relevant in the case of couplings to light fermions. To account for this, we include contributions from $\langle \sigma v \rangle_{\gamma\gamma}$ and $\langle \sigma v \rangle_{\bar{f}f\gamma}$ channels (and $\langle \sigma v \rangle_{gg}$, $\langle \sigma v \rangle_{\bar{f}fg}$ for colored mediators), using analytic results adapted from~\cite{garny2012dark,Garny:2015wea} for Majorana DM and~\cite{toma2013internal,Ibarra:2014qma} for scalar DM. For models with DM coupling to left-handed lepton doublet $\psi=\ell_L$, also the channel $\langle\sigma v\rangle_{W\ell\bar{\nu}}$ is included.
We neglect non-perturbative effects like the Sommerfeld enhancement~\cite{sommerfeld1931beugung} or bound state formation in the calculation, which can be important in the coannihilation regime~\cite{garny2022bound,becker2022impact,biondini2019scalar,Becker:2024vyd}.

Using \texttt{micrOMEGAs}~\cite{Belanger:2006is}, we compute the relic abundance for each model. The parameter space in which the abundance of DM does not overclose the universe, \emph{i.e.} $\Omega h^2 <0.1200\pm 0.0012$~\cite{Planck:2018vyg,ParticleDataGroup:2018ovx}, is cosmologically allowed and, as such, constitutes the target region for (in)direct detection experiments and a Tera-$Z$ factory.

\subsection{Direct Detection} 
\label{sec:DD}

The specific DD phenomenology depends primarily on the spin of the dark matter candidate and the color charge of the mediator. In the following, we will outline the main features, but details on the calculations can be found in App.~\ref{app:DD}.

For both a real scalar and a Majorana fermion DM candidate, vector current interactions with quarks vanish identically due to the properties of the fields involved. As a result, the dominant loop-induced contribution arises from the Higgs penguin diagram, which generates scalar interactions with nucleons.\footnote{An exception to this occurs in the case of axial-vector currents mediated by the $Z$ boson, which induce spin-dependent interactions. However, these are typically much less constrained by current direct detection experiments.}

For models with colored mediators coupled to light quarks—whether involving scalar or fermionic dark matter—the $t$-channel annihilation mechanism implies an $s$-channel scattering of dark matter with quarks, leading to spin-independent (SI) interactions with nucleons. This process constitutes the dominant contribution to direct detection, especially in the regime where the dark matter and mediator masses are nearly degenerate ($M_{\mathrm{DM}} \simeq M_{\mathrm{Med}}$). In addition to the tree-level contribution, loop-induced effects, such as Higgs penguin diagrams and box diagrams involving quark or gluon exchange, can also play a significant role.

In contrast, leptophilic models with uncolored mediators do not give rise to tree-level DM–nucleon scattering. In these cases, the only contribution to SI scattering arises from a loop-induced Higgs penguin diagram, which is further suppressed by the small lepton masses.

We compute the spin-independent and spin-dependent (SD) cross sections for each model using \texttt{Package-X}~\cite{Patel:2015tea}, incorporating partial results from Refs.~\cite{Gondolo:2013wwa,Hisano:2015bma,Mohan:2019zrk,Arcadi:2023imv}. Experimental constraints are then interpreted using \texttt{micrOMEGAs}~\cite{Belanger:2006is}, through which we extract the 90\% CL exclusion limits using results from LZ5T~\cite{LZ:2022lsv}, XENON1T~\cite{XENON:2018voc,XENON:2019rxp} and PICO-60~\cite{PICO:2019vsc} for our mass region of interest. For reference, we also compute the future limits set by the projected sensitivity of DARWIN \cite{DARWIN:2016hyl,Schumann:2015cpa}.

\subsection{Indirect Detection}
\label{sec:indirect_det}

Dark matter self-annihilations can lead to an excess of cosmic-ray fluxes reaching the Earth. The most stringent limits from such indirect detection (ID) arise from $\gamma$-ray observations of dwarf spheroidal galaxies, searches for monochromatic $\gamma$ lines from the Galactic center, and antiproton flux measurements~\cite{Garny:2015wea, Garny:2018icg,Arina:2025zpi}.

Today, the DM annihilation cross-section is dominated by the same channels that set its relic abundance, described in Sec.~\ref{sec:relic_abundance}. In addition, for the spectrum of photons and antiprotons, the loop-suppressed channels $\mathrm{DM}\;\mathrm{DM}\to \gamma \gamma$ and $\mathrm{DM}\;\mathrm{DM}\to g g$ can be important.  The relative importance of $\langle\sigma v\rangle_{\gamma\gamma(gg)}$ with respect to $\langle\sigma v\rangle_{\bar{f}f\gamma(g)}$ is controlled by the ratio $M_{\mathrm{Med}}/M_{\mathrm{DM}}$. 

To derive ID constraints, we first consider $\gamma$-ray observations of dwarf spheroidal galaxies. Using \texttt{micrOMEGAs}, we compute the continuous $\gamma$-ray flux by integrating the photon annihilation spectrum over the line of sight and detector opening angle for each source. We then integrate this flux over each energy bin to obtain the photon energy flux, which we compare with the binned likelihoods published by Fermi-LAT~\cite{Fermi-LAT:2016uux}. We include the nine dwarf galaxies with the largest $J$-factors~\cite{Garny:2018icg,Geringer-Sameth:2014yza} and construct the total log-likelihood by summing over all bins and sources.

Additional constraints arise from searches for monochromatic $\gamma$-ray lines, which are particularly sensitive to loop-induced annihilations into photons. In our models, the process $\mathrm{DM}\,\mathrm{DM} \to \gamma\gamma$ yields two photons with $E_\gamma = M_{\mathrm{DM}}$, producing a sharp spectral feature distinguishable from astrophysical backgrounds~\cite{Bergstrom:1989jr,Flores:1989ru,Srednicki:1985sf,Bergstrom:1988fp}.
Furthermore, internal bremsstrahlung can generate a similar spectral shape, especially when the dark matter and mediator masses are nearly degenerate, and the two signals cannot be disentangled with current energy resolution~\cite{Bergstrom:2004cy,Bringmann:2007nk,Bergstrom:2010gh,Okada_2015}. Thus, the differential $\gamma$-ray flux at Earth is given by~\cite{Garny:2013ama}
\begin{align}
    \frac{d\Phi}{dE_\gamma d\Omega}=&\frac{1}{4\pi}\Big(
    \frac{d\langle\sigma v\rangle_{\bar{f}f\gamma}}{dE_{\gamma}} + 2\langle\sigma v\rangle_{\gamma\gamma}\delta(E_\gamma-M_{\mathrm{DM}})
    \Big)\nonumber\\
    &\times\int_0^{\infty} ds \,\frac{1}{2}\Big(\frac{\rho_{\mathrm{DM}}}{M_{\mathrm{DM}}}\Big)^2,
\end{align}
where we assume a radial dark matter distribution $\rho_{\mathrm{DM}}(r(s))$ given by the Einasto profile \cite{Einasto:1965czb}. We then integrate it for the energy resolution at H.E.S.S.~\cite{HESS:2018cbt} and confront the prediction for each model with the 95\% CL limits set in \cite{HESS:2018cbt}.
Analytic expressions for $\langle\sigma v\rangle_{\gamma\gamma},\langle\sigma v\rangle_{\bar{f}f\gamma}$ can be found in ~\cite{garny2012dark,Garny:2015wea} for Majorana DM and ~\cite{toma2013internal,Ibarra:2014qma} for scalar DM. For comparison, we also compute the limits set by the future CTAO projected sensitivity to line searches, taken from Ref.~\cite{CTAO:2024wvb}.

Finally, the portals in Tab.~\ref{table:models} are also subject to constraints from cosmic-ray antiproton data measured by AMS-02~\cite{AMS:2016oqu}, especially in scenarios with colored mediators. We compute the annihilation cross section $\langle\sigma v\rangle$ into the relevant primary channels using \texttt{micrOMEGAs}, including the loop-induced contribution to gluon final states from~\cite{Garny:2015wea}. The resulting fluxes are then compared to the 95\% CL upper limits derived in~\cite{Cuoco:2017iax} for each annihilation channel.

\section{Results}
\label{sec:results}

\subsection{Fermionic DM}
\label{sec:fermionicDM}

For fermionic dark matter, five portal interactions are possible, corresponding to couplings with each type of SM fermion. Allowing couplings to arbitrary SM flavours generically leads to unsuppressed flavour violation at one-loop, effectively pushing the new physics scale beyond the reach of future Tera-$Z$ factories. The observed suppression of flavour-changing processes requires a non-generic flavour structure if new physics is to appear at the TeV scale, as suggested by Higgs naturalness.

A minimal and experimentally viable scenario consistent with this expectation is one in which NP couples predominantly to third-generation quarks (see also~\cite{Demetriou:2025ewa}). As we show, such couplings can still induce sizable effects in electroweak and cosmological observables, making this scenario a promising target for future $e^+e^-$ facilities. In contrast, NP coupled mainly to third-generation leptons is less accessible at Tera-$Z$, motivating the exploration of alternative flavour structures that remain compatible with current low-energy constraints. 

In the quark-only scenario, mediators carrying QCD charge are constrained by LHC searches due to their pair production via strong interactions. Reinterpretations of existing analyses~\cite{CMS:2019zmd,ATLAS:2020xgt,ATLAS:2020syg,ATLAS:2021kxv,CMS:2021far} in Ref.~\cite{Arina:2025zpi} exclude mediator masses up to $M_\Phi = 0.8$ TeV ($M_\Phi = 1.1$ TeV) for couplings to $\psi = u_R$ ($\psi = q_L$) assuming a DM mass below $0.4$ TeV. Since we focus on the regime $M_{\rm DM}>0.5$ TeV, these bounds do not constrain our parameter space for scalar mediators coupled to quarks. Modest improvements are expected at the High-Luminosity (HL) LHC~\cite{CidVidal:2018eel,ATLAS:2013jyn,ATLAS:2014jim,TheATLAScollaboration:2014nwe,Ruhr:2016xsg}, potentially covering $M_{\rm DM}<0.7$ TeV, though still far from the projected reach of the Tera-$Z$ factory. In the leptonic case, the constraints are even weaker due to suppressed production cross sections for the $t$-channel mediators.

In the following, we focus on $t$-channel portals to quarks with $\psi = \{u_R, q_L\}$, omitting the case $\psi=d_R$. The reason is that in this case, the running effects—crucial for Tera-$Z$ sensitivity as we will see—are suppressed by the bottom Yukawa coupling. On the other hand, in the case of portals to leptons, there are already sizable contributions at the matching scale to the operators $O_{He} = (H^\dagger i\overset{\text{\scriptsize$\leftrightarrow$}}{D}_\mu H)(\bar{e}_R \gamma^\mu e_R)$, $O_{H\ell}^{(1)} = (H^\dagger i\overset{\text{\scriptsize$\leftrightarrow$}}{D}_\mu H)(\bar{\ell}_L \gamma^\mu \ell_L)$, and its $\mathrm{SU}(2)_L$-triplet counterpart $O_{H\ell}^{(3)} = (H^\dagger i\overleftrightarrow{D}_\mu^I H)(\bar{\ell}_L \sigma^I \gamma^\mu \ell_L)$~\cite{Grzadkowski:2010es}. These operators modify the $Z$-boson couplings to leptons and will be probed with high precision, enabling relevant constraints on dark matter models with $t$-channel portals to leptons.

\subsubsection{\texorpdfstring{$\chi\, \Phi\,q_L$}{chi-Phi-qL}}
\label{subsec:FSq}

Let us start with a colored scalar mediator $\Phi \sim (\mathbf{3},\mathbf{2},1/6)$ connecting the dark matter candidate to the third-generation quark doublet. The relevant Lagrangian terms are
\begin{align}
 -\mathcal{L} &\supset \kappa_1 |\Phi|^2 |H|^2 
 +\kappa_2(\Phi^\dagger \sigma^I \Phi) (H^\dagger \sigma^I H)\nonumber\\
 &\quad + \left(y_Q \overline{\chi} q_L^3 \Phi^\dagger + \mathrm{h.c.}\right)\,.
\end{align}
Having a heavy dark sector, we perform the matching onto the SMEFT using \texttt{matchmakereft}~\cite{Carmona:2021xtq}, and study the parameter space spanned by $y_Q$, $\kappa_1$, $\kappa_2$, and the dark matter and the mediator masses $M_\chi$ and $M_\Phi$, respectively.

\begin{figure}[t]
    \centering
    \vspace{-0.15cm}\includegraphics[width=\linewidth]{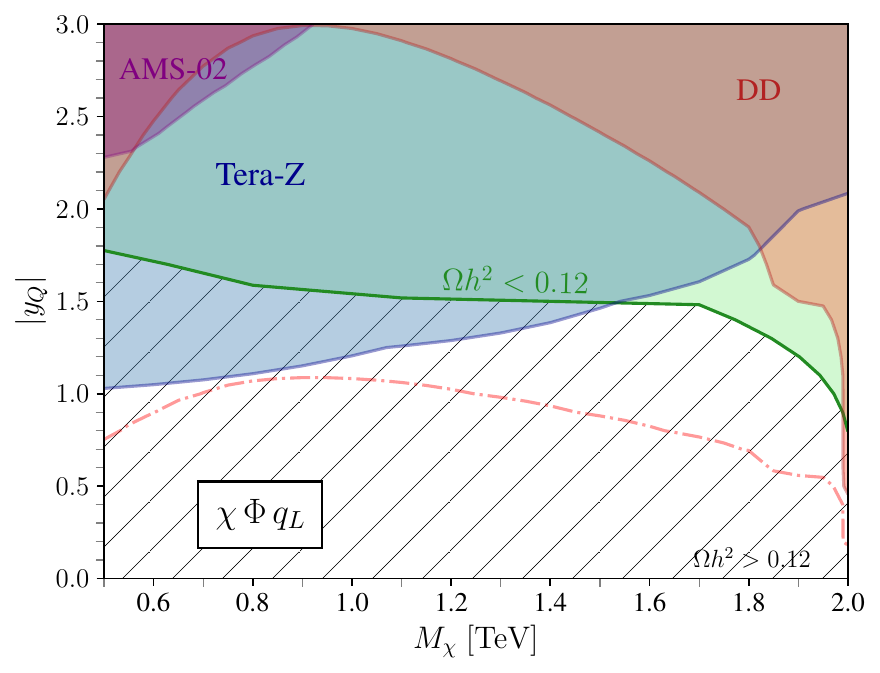}
     \includegraphics[width=\linewidth]{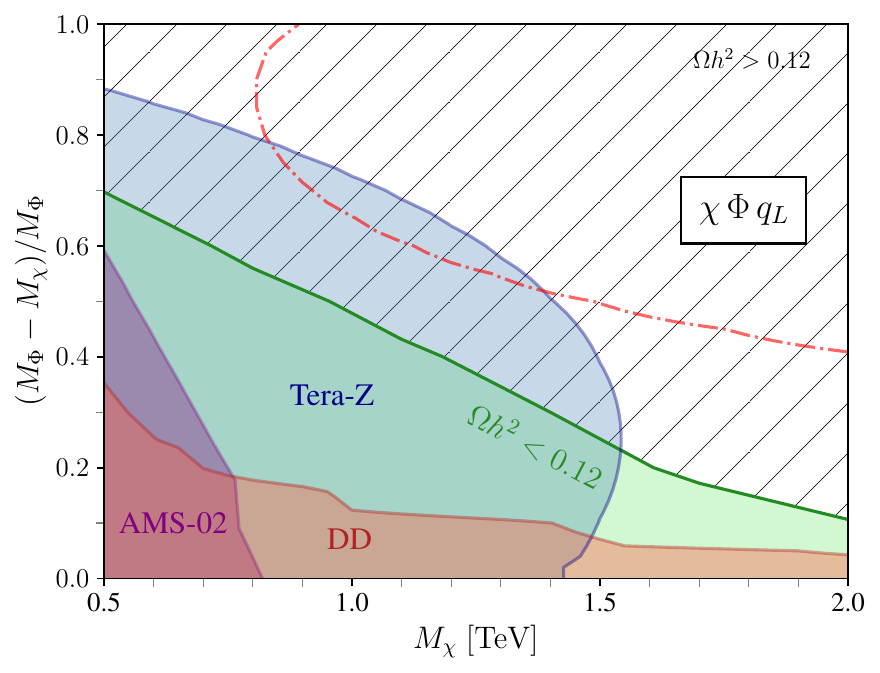}
    \caption{\emph{$\chi \Phi q_L$ case:} Parameter space with fixed mediator mass $M_\Phi = 2$ TeV (upper panel) and fixed portal coupling $y_Q=1.5$ (lower panel). The solid green line indicates the parameter values yielding the observed dark matter relic abundance. The surrounding green region corresponds to scenarios where $\chi$  is a subcomponent of dark matter and represents the \emph{target} parameter space. The blue region is excluded by the projected Tera-$Z$ sensitivity at 95\% CL for $\kappa_{1,2}=0$ while current direct detection bounds exclude the red region. The DARWIN projections exclude the parameter space below the dash-dotted red line.}
    \label{fig:Fs_Sql}
\end{figure}

In Fig.~\ref{fig:Fs_Sql} we show the parameter space with fixed mediator mass $M_\Phi = 2$ TeV (upper panel) and fixed portal coupling $y_Q=1.5$ (lower panel). The solid green line indicates the parameter values yielding the observed dark matter relic abundance. The surrounding green region corresponds to scenarios where $\chi$ is a subcomponent of dark matter and represents the \emph{target} parameter space. The blue region is excluded by the projected sensitivity at a future Tera-$Z$ run at 95\% CL for $\kappa_{1,2}=0$, while current direct detection bounds exclude the red region. The DARWIN~\cite{DARWIN:2016hyl} projections exclude the parameter space below the dash-dotted red line.

A remarkable complementarity emerges between the Tera-$Z$ sensitivity and the region where $\chi$ can account for the dark matter relic abundance. Even though the DARWIN experiment is expected to fully cover the parameter space compatible with thermal freeze-out, a potential Tera-$Z$ factory would play a decisive role in interpreting a possible  signal. As we highlight below, correlations among EWPOs can provide detailed information on the underlying NP scenario.

\begin{figure}[t]
    \centering
    \includegraphics[width=\linewidth]{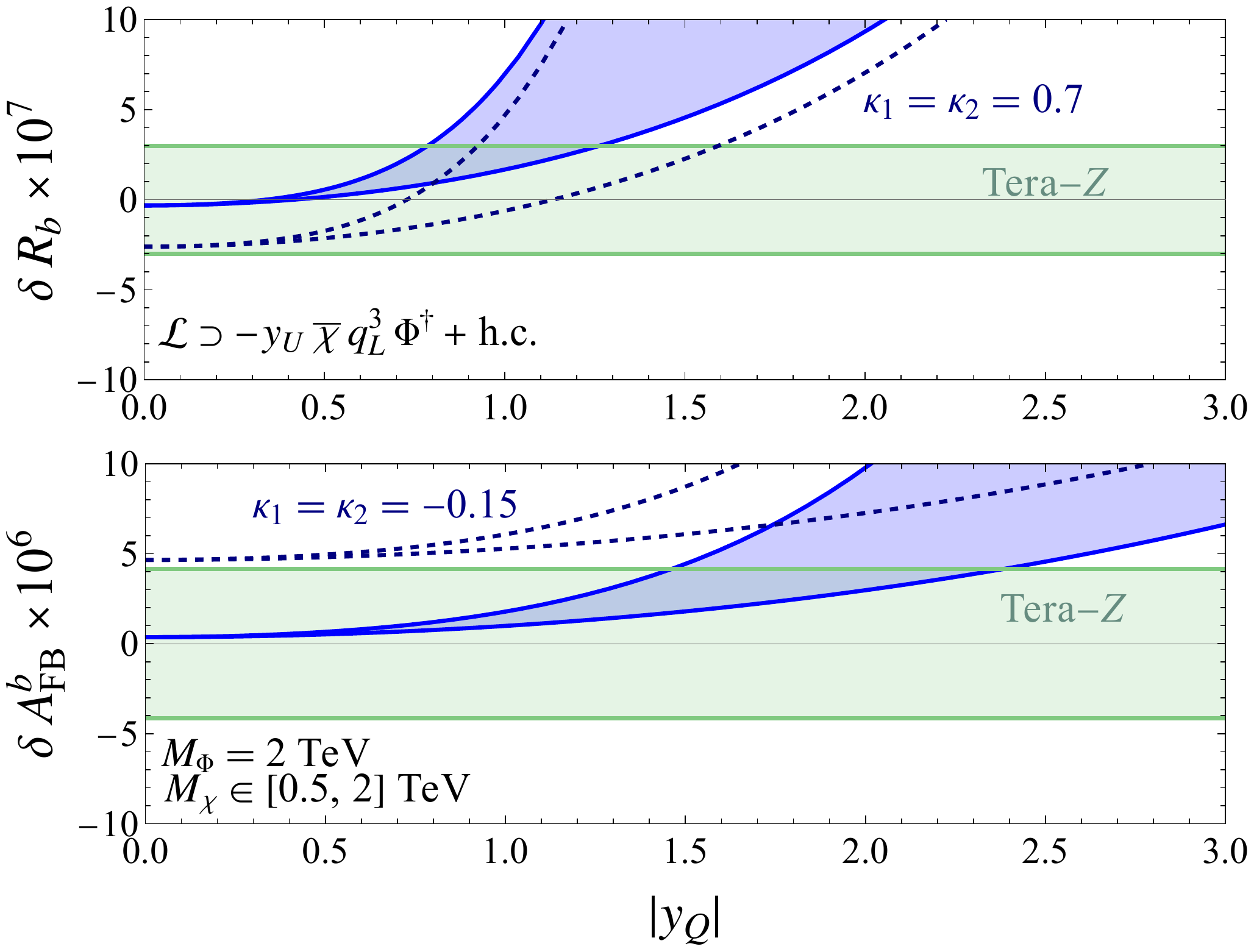}
    \caption{\emph{$\chi \Phi q_L$ case:} The variation of the most influential electroweak observables, $R_b$ and $A_{\rm FB}^b$, as a function of the portal coupling. The green bands denote the Tera-$Z$ experimental sensitivity at $1\sigma$, while the blue bands reflect the variation of the dark matter mass $M_\chi \in [0.5, 2]$ TeV. The mediator mass is fixed to $M_\Phi = 2$ TeV. The dashed lines show the effect of non-vanishing $\kappa_1$ and $\kappa_2$, where the main effect comes from $\kappa_2$, see the main text.}
    \label{fig:FSq_EWPOs}
\end{figure}

In particular, the observables with the largest pull, $P_O$ defined in Eq.~\eqref{eq:pulls}, in this scenario are $R_b$ and $A_{\rm FB}^b$. Their dependence on the portal coupling is illustrated in Fig.~\ref{fig:FSq_EWPOs}. The green bands denote the $1\sigma$ experimental sensitivity, while the blue bands reflect the variation of the dark matter mass $M_\chi \in [0.5, 2]$ TeV, keeping the mediator mass fixed to $M_\Phi = 2$ TeV. One can appreciate that already $R_b$ (at $2\sigma$) reflects the bounds on the portal coupling shown in Fig.~\ref{fig:Fs_Sql} (upper panel) to a good extent. 

\begin{figure}[t]
    \centering
    \includegraphics[width=\linewidth]{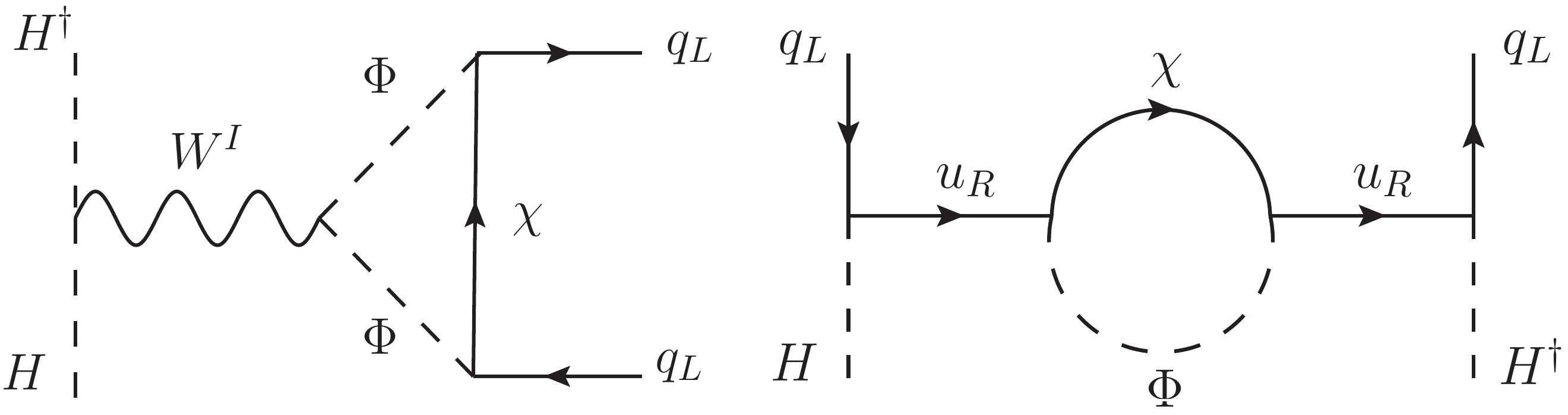}
    \caption{Feynman diagrams showing the leading-order contributions to the electroweak precision observables in the $\chi \Phi q_L$ scenario (left) and the $\chi \Phi u_R$ scenario (right).}
    \label{fig:Feyn_diagrams}
\end{figure}

From the EFT perspective, this happens due to the shift in the $Z b_L b_L$ vertex which is proportional to $[C_{Hq}^{(1)}+C_{Hq}^{(3)}]_{33}$ and is mainly generated through the left diagram shown in Fig.~\ref{fig:Feyn_diagrams}. First, a contribution to the redundant operators $O_{Bq} = (\bar{q}_L\gamma^\mu q_L)\partial^\nu B_{\mu\nu}$ and $O_{Wq} = (\bar{q}_L\sigma^I \gamma^\mu q_L) D^\nu W_{\mu\nu}^I$ in the Green's basis is generated~\cite{Gherardi:2020det}, after which the rotation to the Warsaw basis operators $O_{Hq}^{(1,3)}$ has been performed~\cite{Grzadkowski:2010es}. 

The combination $[C_{Hq}^{(1)}+C_{Hq}^{(3)}]_{33}$ is dominated by the $g_2$ contributions to $[C_{Hq}^{(3)}]_{33}$, where $g_2$ is the $\mathrm{SU}(2)_L$ coupling, and it reads
\begin{equation}
    [C_{Hq}^{(3)}]_{33} \simeq \frac{g_2^2 |y_Q|^2}{2304\pi^2 M_\chi^2} 
    \!\left[\!\frac{x(2x-7)+11}{(x-1)^3}-\frac{6\log(x)}{(x-1)^4}\!\right]\!,
\end{equation}
with $x=M_\Phi^2/M_\chi^2$. Moreover, it is always positive for $x>1$, and entering positively both $R_b$ and $A_{\rm FB}^b$, resulting in positive predictions in Fig.~\ref{fig:FSq_EWPOs}. 

\begin{figure}[t]
    \centering
    \includegraphics[width=0.85\linewidth]{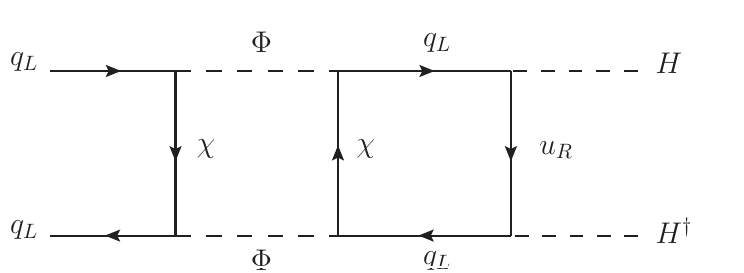}
    \caption{Feynman diagram showing the next-to-leading-order contribution to the electroweak precision observables in the $\chi \Phi q_L$ case.}
    \label{fig:2loop}
\end{figure}

In addition, we find that a two-loop effect, illustrated in Fig.~\ref{fig:2loop}, significantly enhances the sensitivity of Tera-$Z$. This effect arises from sizable one-loop contributions to the four-quark operators $O_{qq}^{(1)} = (\bar{q}_L \gamma^\mu q_L)(\bar{q}_L \gamma_\mu q_L)$ and $O_{qq}^{(3)} = (\bar{q}_L \sigma^I \gamma^\mu q_L)(\bar{q}_L \sigma^I \gamma_\mu q_L)$ involving third-generation quark doublets. These operators subsequently mix into $O_{Hq}^{(1,3)}$ through renormalization group evolution, driven by a large anomalous dimension proportional to the top-Yukawa coupling~\cite{Jenkins:2013wua}. At leading-log order, this yields
\begin{align}
    &C_{Hq}^{(1)}+C_{Hq}^{(3)}\simeq \frac{4N_c |y_t|^2}{16\pi^2}\!\left[C_{qq}^{(1)}-\frac{2N_c-1}{2N_c}C_{qq}^{(3)}\!\right]\!\log\!\left(\!\frac{\mu}{M_\chi}\!\right)\nonumber\\
    =&\,\frac{3|y_Q|^4 |y_t|^2}{512\pi^4 M_\chi^2}\!\left[\!\frac{x+5}{(x-1)^2}-\frac{(4x+2)\log(x)}{(x-1)^3}\!\right]\!\log\!\left(\!\!\frac{\mu}{M_\chi}\!\right)\!,\!
    \label{eq:2loop-FSq}
\end{align}
where the flavour indices are third-generation implicit, $N_c=3$ is the number of colors, and $\mu$ corresponds to the electroweak scale at which the observables are measured. Similarly to $C_{Hq}^{(3)}$, it is always positive and adds to the total sensitivity. 

Furthermore, we employed the evolution matrix method implemented in \texttt{DSixTools}~\cite{Celis:2017hod} to express the EWPOs at the low scale $\mu \approx m_Z$ in terms of the Wilson coefficients defined at the UV matching scale. This approach allows us to consistently incorporate the full renormalization group evolution, capturing all running effects. Our numerical analysis confirms that the two-loop contribution in Eq.~\eqref{eq:2loop-FSq} is the most prominent one, leading to a significantly stronger constraint on $|y_Q|$. Specifically, including this contribution improves the bound up to a factor of two compared to when it is neglected.

Due to the mass splitting between the dark matter candidate and the mediator, the technically correct procedure would be to integrate out the mediator at its mass scale and run the resulting Wilson coefficients down to $M_{\mathrm{DM}}$ scale.
However, in our setup, the hierarchy is never large, with typical corrections of at most $\log(M_{\rm Med}/M_{\rm DM})\simeq 1.4$, rendering the effect numerically small. Moreover, since we apply one-loop RGEs on top of the finite one-loop matching, our result is formally two-loop order. A full two-loop matching computation is beyond the scope of this work, and would only lead to slight corrections to our predictions.

Next, we comment on the impact of the quartic couplings of the scalar mediator to the Higgs boson. Most prominently, they affect the operator $O_{HWB} = (H^\dagger \sigma^I H) W^I_{\mu\nu} B^{\mu\nu}$, which affects the $Z$ boson mass that is an input parameter and thus enters universally into all electroweak processes. In the case of the $\chi \Phi q_L$ portal, we have
\begin{equation}
    C_{HWB} = \frac{g_1 g_2\kappa_2}{192 \pi^2 M_{\Phi}^2}\,,
    \label{eq:HWB}
\end{equation}
and we find $A_{\rm FB}^b$ to be particularly sensitive to it. This is illustrated in the lower panel of Fig.~\ref{fig:FSq_EWPOs}, where the dashed lines show that already this observable constrains $|\kappa_{2}|\lesssim 0.15$ for the mediator mass of $M_\Phi = 2$ TeV.


\begin{figure}[t]
    \centering
    \includegraphics[width=\linewidth]{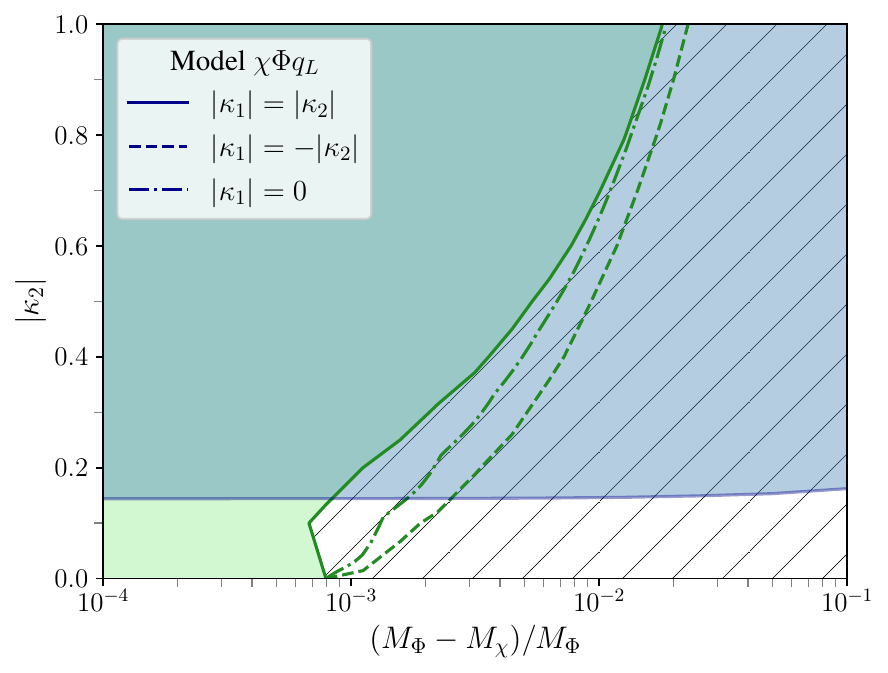}
    \caption{\emph{$\chi\Phi q_L$ case:} Parameter space with fixed DM mass $M_\chi=1.5$ TeV and $y_Q=0$. The green lines indicate the values reproducing the observed relic density for different values of $\kappa_1$. The blue lines represent the exclusion limits by the projected Tera-$Z$ sensitivity at 95\% CL.}
    \label{fig:kappa}
\end{figure}

This effect further highlights the complementarity of a Tera-$Z$ machine with direct and indirect dark matter searches. Indeed, in the coannihilation regime, the couplings $\kappa_{1,2}$ open a Higgs portal to SM, allowing the observed relic density to be achieved even in the limit $y_Q = 0$. However, this scenario is challenging to probe since $\kappa_{1,2}$ only mediate DM-annihilation or scattering with nuclei proportional to $y_Q$. A Tera-$Z$ machine would offer \emph{unique} sensitivity in this regime, capable of excluding couplings $\kappa_{2}\sim\mathcal{O}(0.1)$ couplings in the $M_{\Phi}\sim\mathcal{O}(1)$ TeV range - parameter space otherwise inaccessible to both direct and indirect detection strategies.
In Fig.~\ref{fig:kappa} we represent the target region of relic density as a function of $\kappa_2$ and the mass splitting for $M_\chi=1.5$ TeV and $y_Q=0$ for different values of $\kappa_1$. 

Finally, the spin-independent scattering with nucleons in this model is mediated by the Higgs penguin diagram, which generates a scalar coefficient $c_S$ (see Eq.~\eqref{eq:DD_majorana_lag} in App.~\ref{app:DD}) greatly enhanced by the mass of the top quark. Loop contributions to $c_G^{(b)}$, $g_G^{(b)}$ with the bottom quark are also important, particularly in the limit $M_\chi\ll M_\Phi$. This effect, in addition to exceptional sensitivity at DARWIN, makes DD the main future probe of this portal.

\subsubsection{\texorpdfstring{$\chi\,\Phi\,u_R$}{chi-Phi-uR}}
\label{subsec:chi_Phi_u}

Next, we study a colored scalar mediator $\Phi\sim(\mathbf{3},\mathbf{1},2/3)$ connecting the dark matter candidate to the top-right quarks. The relevant Lagrangian terms are
\begin{equation}
-\mathcal{L} \supset \kappa |\Phi|^2 |H|^2 + \left(y_U \overline{\chi} u_R^3 \Phi^\dagger + \mathrm{h.c.}\right)\,.
\end{equation}

\begin{figure}[t]
    \centering
    \vspace{-0.15cm}\includegraphics[width=\linewidth]{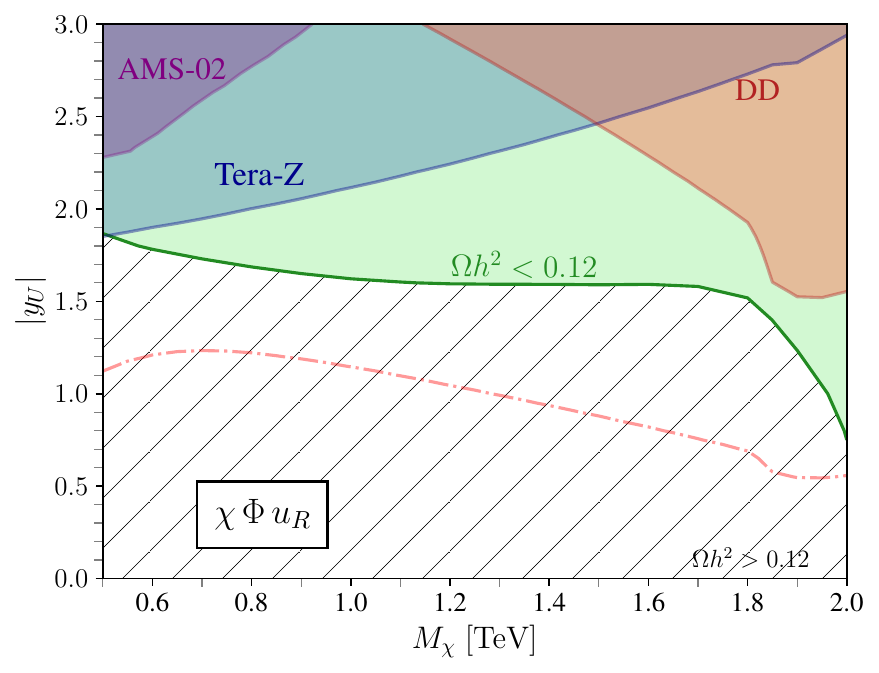}
    \includegraphics[width=\linewidth]{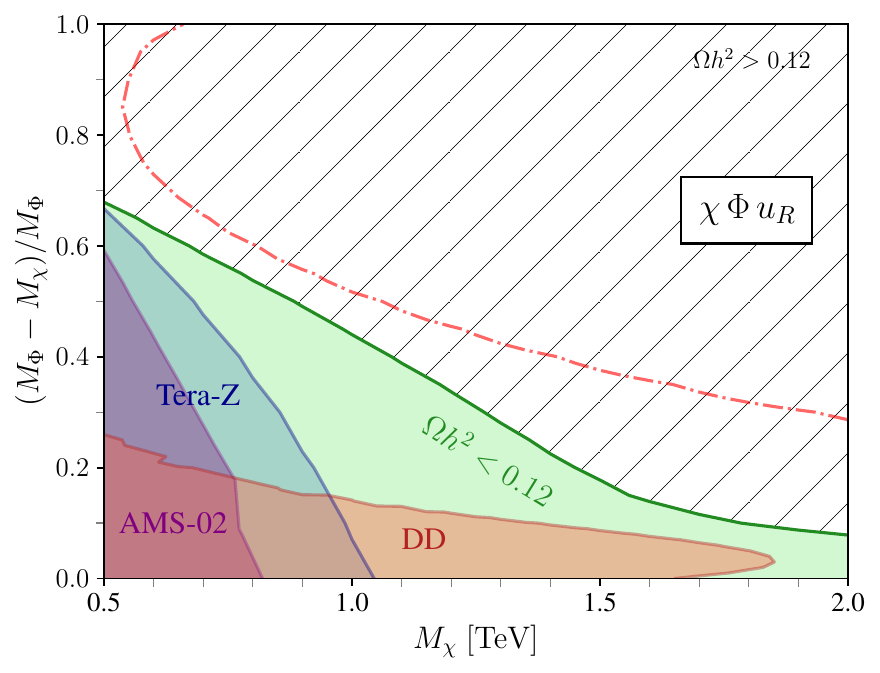}
    \caption{\emph{$\chi \Phi u_R$ case:} Parameter space with fixed mediator mass $M_\Phi = 2$ TeV (upper panel) and fixed portal coupling $y_U=1.5$ (lower panel). Labels and regions same as in Fig.~\ref{fig:Fs_Sql}.}
    \label{fig:chi_Phi_u}
\end{figure}
In Fig.~\ref{fig:chi_Phi_u} we show the parameter space analogous to the $\chi\Phi q_L$ portal in Fig.~\ref{fig:Fs_Sql}. The observables with the largest pull are also $R_b$ and $A_{\rm FB}^b$, demonstrating their exceptional power in constraining new physics coupling to third-generation quarks.
\begin{figure}[t]
    \centering
    \includegraphics[width=\linewidth]{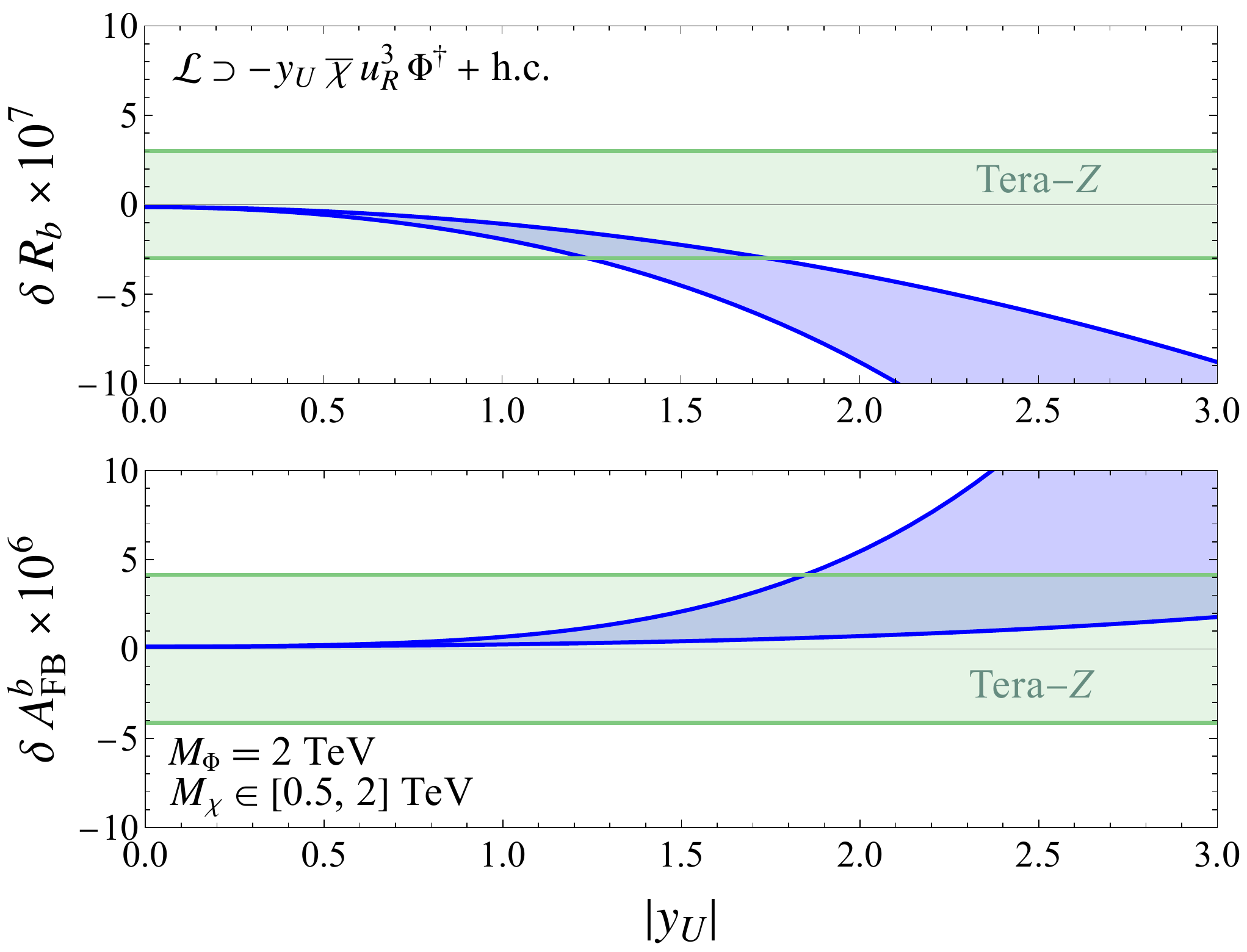}
    \caption{\emph{$\chi \Phi u_R$ case:} The variation of the most influential electroweak observables, $R_b$ and $A_{\rm FB}^b$, as a function of the portal coupling. The green bands denote the Tera-$Z$ experimental sensitivity at $1\sigma$, while the blue bands reflect the variation of the dark matter mass $M_\chi \in [0.5, 2]$ TeV. The mediator mass is fixed to $M_\Phi = 2$ TeV, while $\kappa$ does not influence the EWPOs in this case.}
    \label{fig:FSu_EWPOs}
\end{figure}
In this case, the combination $[C_{Hq}^{(1)}+C_{Hq}^{(3)}]_{33}$ receives contributions from both diagrams in Fig.~\ref{fig:Feyn_diagrams}. The diagram on the right, contributing to $O_{uD} = \frac{i}{2}\bar{u}_R \{D_\mu D^\mu, \slashed{D}\}u_R$, is naively the leading piece because the rotation to $O_{Hq}^{(1,3)}$ involves the insertion of the top-Yukawa coupling, and $|y_t|^2 \gg g_1^2$, with $g_1$ being the ${\rm U(1)}_Y$ gauge coupling. However, the contributions to $C_{Hq}^{(1)}$ and $C_{Hq}^{(3)}$ cancel precisely, leaving
\begin{equation}
   \left[C_{Hq}^{(1)}+C_{Hq}^{(3)}\right]_{33} = - \frac{g_1^4}{4320 \pi^2 M_{\Phi}^2}\,.
   \label{eq:CHq1+3_FSu}
\end{equation}
If this were the only contribution, the indirect signals of the $\chi \Phi u_R$ scenario at the future $e^+e^-$ facilities would be unattainable. Nevertheless, there is again an important two-loop effect stemming from the large self-anomalous dimensions of the $O_{Hq}^{(1,3)}$ operators. Indeed, at the leading-log
\begin{align}
 &\left[C_{Hq}^{(1)}+C_{Hq}^{(3)}\right]_{33} \simeq -\frac{11 |y_t|^2}{32\pi^2} \left[C_{Hq}^{(3)}\right]_{33}\log\!\left(\!\frac{\mu}{M_\chi}\!\right)\label{eq:2loop-FSu}\\
    =&\,\frac{11|y_U|^2 |y_t|^4}{12288\pi^4 M_\chi^2}\frac{x^2(x-6)+3x(1+\log x^2)+2}{(x-1)^4}\log\!\left(\!\!\frac{\mu}{M_\chi}\!\right)\!,\!\nonumber
\end{align}
which completely dominates over Eq.~\eqref{eq:CHq1+3_FSu}, and exceeds it by roughly a factor of 30 for the benchmark values $x=2$ and $y_U=1$.  In contrast to the quark doublet portal in Eq.~\eqref{eq:2loop-FSq}, this contribution is always negative, predicting a decrease in $R_b$ with respect to the SM. This distinction offers a handle to discriminate between $t$-channel DM portals to quark doublets versus singlets. Once again, we emphasize the power of Tera-$Z$ runs in probing the indirect NP contributions, opening a new domain in which the \emph{two}-loop matching and running effects bear important phenomenological consequences.

In this model, the dominant contribution to SI scattering once again arises from the $c_S^q$ coefficient enhanced by the mass of the top quark. As a result, DARWIN imposes again particularly stringent constraints.

\subsubsection{\texorpdfstring{$\chi\, \Phi\,e_R$}{chi-Phi-eR}}
\label{subsec:chi_Phi_e}

Since lepton flavour is a symmetry of the SM and charged lepton flavour violation (cLFV) is strongly constrained, we assume that NP also respects individual charged lepton numbers. When the new states couple to right-handed charged leptons, the Tera-$Z$ phenomenology is dominated by the coupling to electrons. Thus, we focus on interactions with the first generation, given by
\begin{equation}
\label{eq:lag_FSe}
-\mathcal{L}\supset \kappa |\Phi|^2 |H|^2 +\left(y_E \overline{\chi}e_R^1 \Phi^\dagger +\mathrm{h.c.}\right)\,.
\end{equation}
This setup captures the main features of more general flavour structures, as long as they do not induce observable cLFV. Moreover, mediators that couple only to leptons are typically unconstrained by LHC direct searches, and, since they mediate interactions with nuclei suppressed by the mass of the leptons, also evade direct detection bounds.

\begin{figure}[t]
    \centering
    \vspace{-0.1cm}\includegraphics[width=\linewidth]{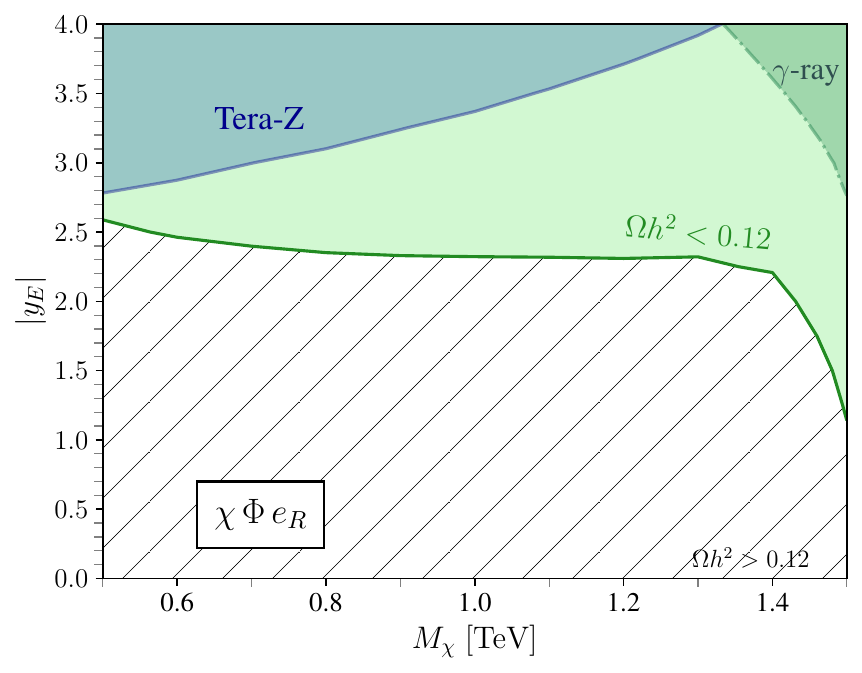}
    \includegraphics[width=\linewidth]{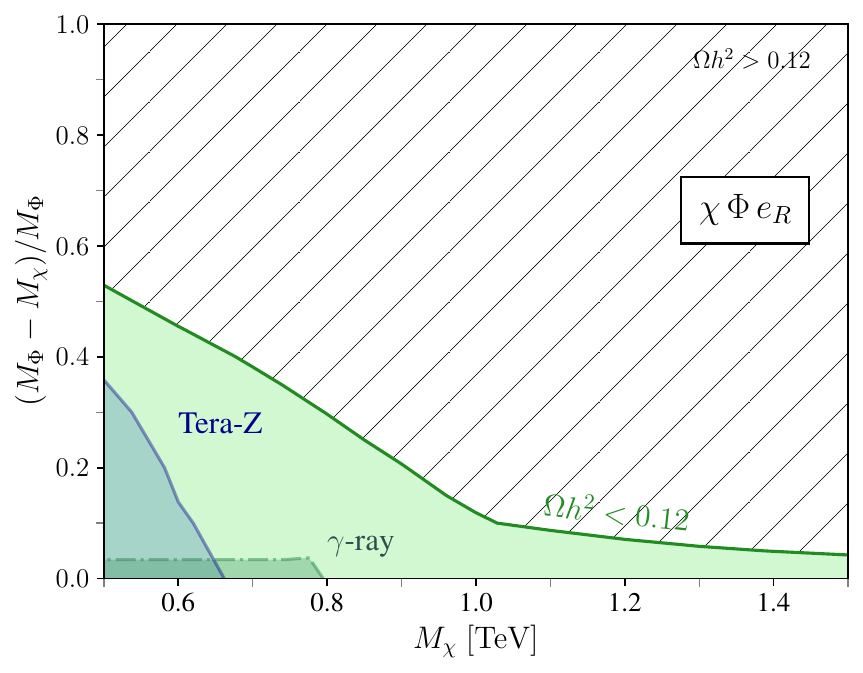}
    \caption{\emph{$\chi \Phi e_R$ case:} Parameter space with fixed mediator mass $M_\Phi = 1.5$ TeV (upper panel) and fixed portal coupling $y_E=2$ (lower panel). Labels and regions same as in Fig.~\ref{fig:Fs_Sql}. Additionally, the dashed green region corresponds to the projected sensitivity of CTAO~\cite{CTAO:2024wvb}.}
    \label{fig:chi_Phi_e}
\end{figure}

Repeating the analysis carried out for the quark portal, we present the results for the $\chi \Phi e_R$ scenario in Fig.~\ref{fig:chi_Phi_e}. In the absence of direct detection constraints, a Tera-$Z$ run may offer the \emph{only} opportunity to probe regions of parameter space where $\chi$ can account for a fraction of the dark matter and there is $\mathcal{O}(1)$ mass splitting with the mediator. In the DM-mediator degeneracy limit, we could learn from the potential $\gamma$-rays from DM  annihilation at the future CTAO observatory~\cite{CTAO:2024wvb}.

\begin{figure}[t]
    \centering
    \includegraphics[width=\linewidth]{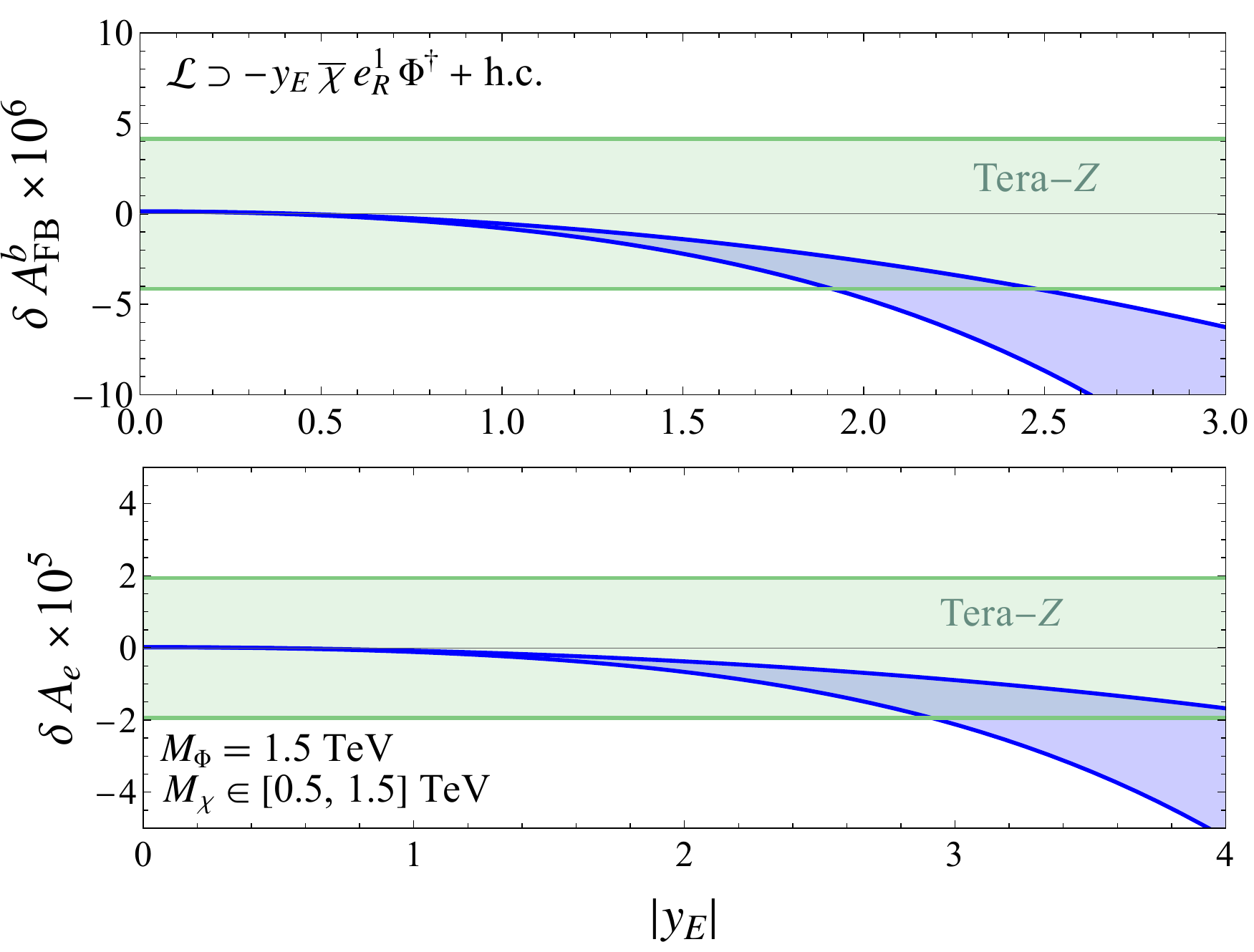}
    \caption{\emph{$\chi \Phi e_R$ case:} The variation of the most influential electroweak observables, $A_{\rm FB}^b$ and $A_e$, as a function of the portal coupling. 
    The green bands denote the Tera-$Z$ experimental sensitivity at $1\sigma$, while the blue bands reflect the variation of the dark matter mass $M_\chi \in [0.5, 1.5]$ TeV. The mediator mass is fixed to $M_\Phi = 1.5$ TeV, while $\kappa$ does not influence the EWPOs in this case.}
    \label{fig:FSe_EWPOs}
\end{figure}

The electroweak observables with the largest pull in this case are $A_{\rm FB}^b$ and the electron left-right asymmetry $A_e$. The NP contribution as a function of the portal coupling $|y_E|$ is shown in Fig.~\ref{fig:FSe_EWPOs}. For this portal, the main effect comes from
\begin{equation}
    [C_{He}]_{11}\! \simeq\! \frac{g_1^2 |y_E|^2}{1152\pi^2 M_\chi^2} 
    \!\left[\!\frac{x(7-2x)-11}{(x-1)^3}+\frac{6\log(x)}{(x-1)^4}\!\right]\!.
    \label{eq:CHe_FSe}
\end{equation}
This contribution is always negative, leading to a correlated decrease in both $A_{\rm FB}^b$ and $A_e$ relative to the SM prediction. Notably, running effects play a relevant role, in particular the Higgs wavefunction renormalization term proportional to $N_c |y_t|^2 [C_{He}]_{11}$. Additional contributions arise from the semileptonic operators $[O_{qe}]_{3311} = (\bar{q}_L^3 \gamma^\mu q_L^3)(\bar{e}_R^1 \gamma^\mu e_R^1)$ and $[O_{eu}]_{1133} = (\bar{e}_R^1 \gamma^\mu e_R^1)(\bar{u}_R^3 \gamma^\mu u_R^3)$, which also run significantly into $[O_{He}]_{11}$, with anomalous dimensions proportional to $\pm 2 N_c |y_t|^2$, respectively~\cite{Jenkins:2013wua}.

Lastly, given that the mediator is an ${\rm SU(2)}_L$ singlet, the quartic coupling $\kappa$ in Eq.~\eqref{eq:lag_FSe} affects only the operators $O_{H}=(H^\dagger H)^3$, $O_{HB}=(H^\dagger H)(B_{\mu\nu}B^{\mu\nu})$, and $O_{H\Box}=(H^\dagger H)\Box(H^\dagger H)$ which have a negligible impact on the EWPOs.

\subsubsection{\texorpdfstring{$\chi\, \Phi\,\ell_L$}{chi-Phi-ellL}}
\label{subsec:chi_Phi_lL}

\begin{figure}[t]
    \centering
    \vspace{-0.1cm}\includegraphics[width=\linewidth]{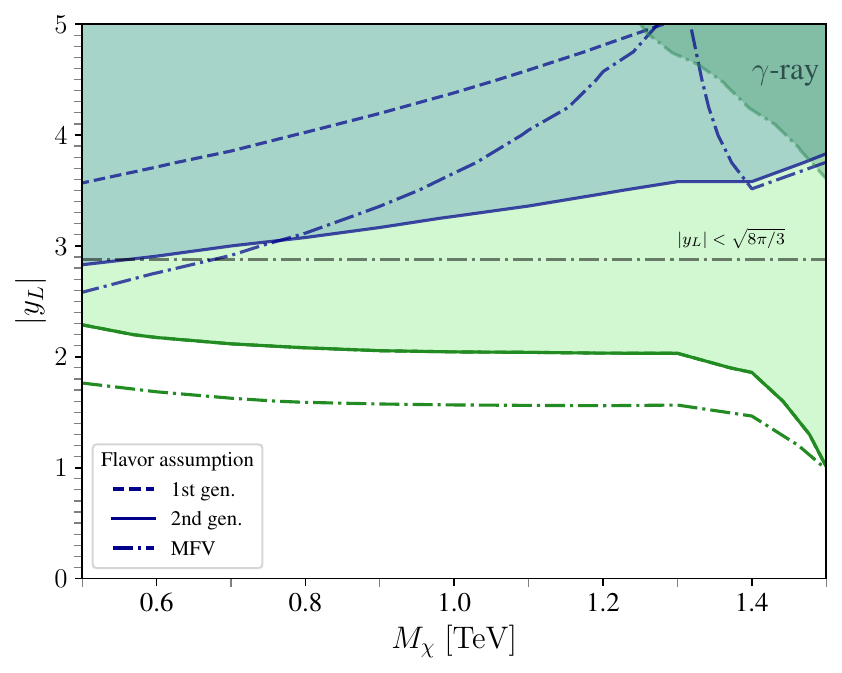}
    \caption{\emph{$\chi \Phi \ell_L$ case}: Parameter space with fixed mediator mass $M_\Phi = 1.5$ TeV for three different flavor assumptions. The solid green lines correspond to the parameter values that yield the observed dark matter relic abundance. The blue lines are the Tera-$Z$ exclusion limits at 95\% CL for $\kappa_{1,2}=0$. The top-right colored corner is excluded by $\gamma$-ray line searches at H.E.S.S. \cite{HESS:2018cbt} for the MFV scenario. The horizontal gray line is the perturbative unitarity limit on the coupling in the case of MFV. } 
    \label{fig:Fs_Sl}
\end{figure}

Next, we study a portal to left-handed lepton doublets, defined by the Lagrangian
\begin{align}
\label{eq:lag_FSell}
    -\mathcal{L}&\supset \kappa_1 |\Phi|^2 |H|^2 +\kappa_2\, (\Phi^\dagger \sigma^I \Phi) (H^\dagger \sigma^I H)\nonumber\\
    &\quad+\left([y_L]_i \overline{\chi}\,\ell_L^i \Phi^\dagger +\mathrm{h.c.}\right)\,,
\end{align}
where $i=\{1,2,3\}$ is the flavour index. In contrast to the previous leptonic model, couplings to second generation are more constrained in this scenario due to the modification to the muon decay induced by $c_{H\ell}\sim g_2^2 y_L^2$. In Fig.~\ref{fig:Fs_Sl}, we analyze the bounds in three different flavor scenarios: couplings only to first generation ($[y_L]_i\propto \delta_{i1}$), second generation ($[y_L]_i\propto \delta_{i2}$), or assuming Minimal Flavor Violation (MFV)~\cite{DAmbrosio:2002vsn}. For this last case, we assumed three degenerate generations for the mediator $\Phi^{j=1,2,3}$
\begin{align}
\label{eq:lag_FSell_MFV}
    -\mathcal{L}&\supset [y_L]_{ij} \overline{\chi}\,\ell_L^i \Phi^{j\dagger} +\mathrm{h.c.}\,,
\end{align}
imposing a $U(3)_\ell$ symmetry on the coupling $[y_L]_{ij}\propto \delta_{ij}$.

The observables with the highest pull for this portal are the forward-backward asymmetry $A_{\mathrm{FB}}^{b}$ and the $Z$-width $\Gamma_Z$, with a correlated decrease in both of them relative to the SM prediction. The main effects come from the coefficients
\begin{align}
    &\frac{1}{g_2^2}[C_{H\ell}]^{(3)}_{ii}=-\frac{1}{g_1^2}[C_{H\ell}]^{(1)}_{ii}=\frac{1}{g_2^2}[C_{\ell\ell}]_{1221}\simeq\nonumber\\
    &\simeq\frac{ |y_L|_i^2}{2304\pi^2 M_\chi^2}\left[\frac{11+ x(7-2x)}{(x-1)^3} -\frac{6\log(x)}{(x-1)^4}\right],
\end{align}
where we assumed a coupling with only one generation $i=\{1,2\}$. In the case of the coupling to first generation, $[C_{H\ell}]^{(1)}_{11}$ enters the asymmetry through the total cross section of $e^+e^-$ to fermions. When coupling to second generation, $[C_{H\ell}]^{(3)}_{22}$ and $[C_{\ell\ell}]_{1221}$ modify the muon decay, that enters $A_{\mathrm{FB}}^{b}$ through $G_F$.

Remarkably, we observe that only in the MFV scenario there is a cancellation stemming from the running of $[C_{\ell\ell}]_{1122}\sim |y_L|^4$ into $[C_{\ell\ell}]_{1221}$, which screens its direct contribution to $m_W$ and $\sigma_{\mathrm{had}}$. As one can see in Fig.~\ref{fig:Fs_Sl}, this significantly weakens the exclusion limits for $M_{\chi}\simeq 1.3$ TeV.
Besides, in the limit of nearly degenerate masses, $\gamma$-ray line searches at H.E.S.S. \cite{HESS:2018cbt} also impose constraints in this scenario. For completeness, we also show the limit on $y_L$ set by partial wave perturbative unitarity. We include the estimates for all models in App.~\ref{sec:th_constraints}.

Finally, having a $\mathrm{SU}(2)_L$ doublet mediator gives rise to a similar phenomenology for the scalar quartic couplings. $C_{HWB}$ is generated with the same size but opposite sign to the one in Eq.~\eqref{eq:HWB}, which constrains $\kappa_2\sim\mathcal{O}(0.1)$.

\subsection{Scalar DM}
\label{subsec:scalar_DM}
When DM is a scalar, there are again five distinct portal interactions corresponding to each SM fermion species. Similarly to the previous cases, generic flavor structures induce loop-level flavor violation that pushes the new physics scale above the multi-TeV regime, unless non-anarchic flavor alignments are assumed.

Compared to the Majorana DM, portals with colored mediators are subject to significantly stronger constraints. In particular, the top-quark–mediated Higgs penguin diagram leads to an enhanced contribution to DM–nucleon scattering, rendering DD bounds especially severe for third-generation couplings. As a result, much of the phenomenologically interesting parameter space is ruled out, motivating the exploration of alternative, data-compatible flavor structures. 

\begin{figure}[t]
    \centering
    \vspace{-0.15cm}\includegraphics[width=\linewidth]{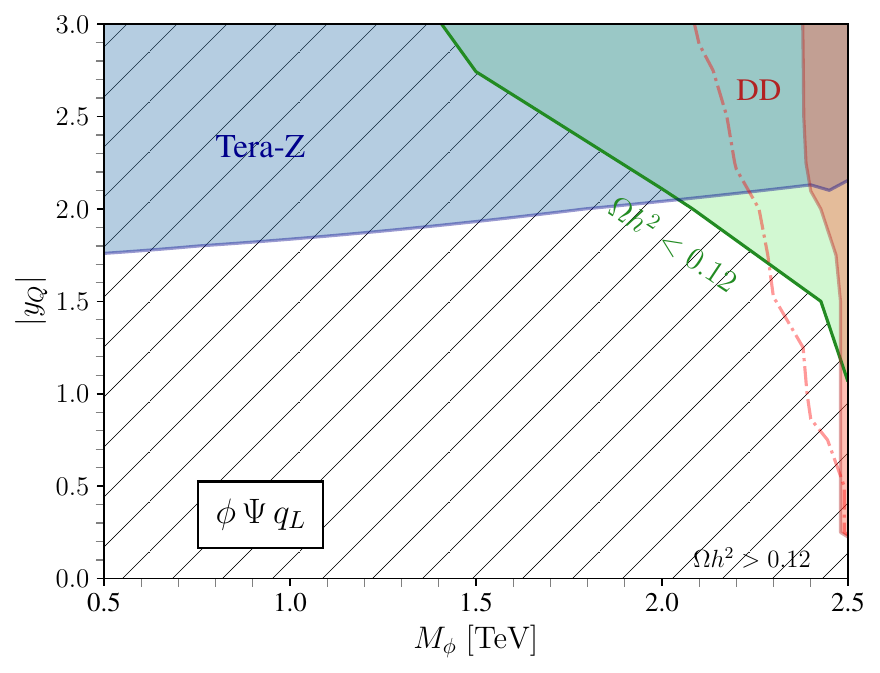}
    \includegraphics[width=\linewidth]{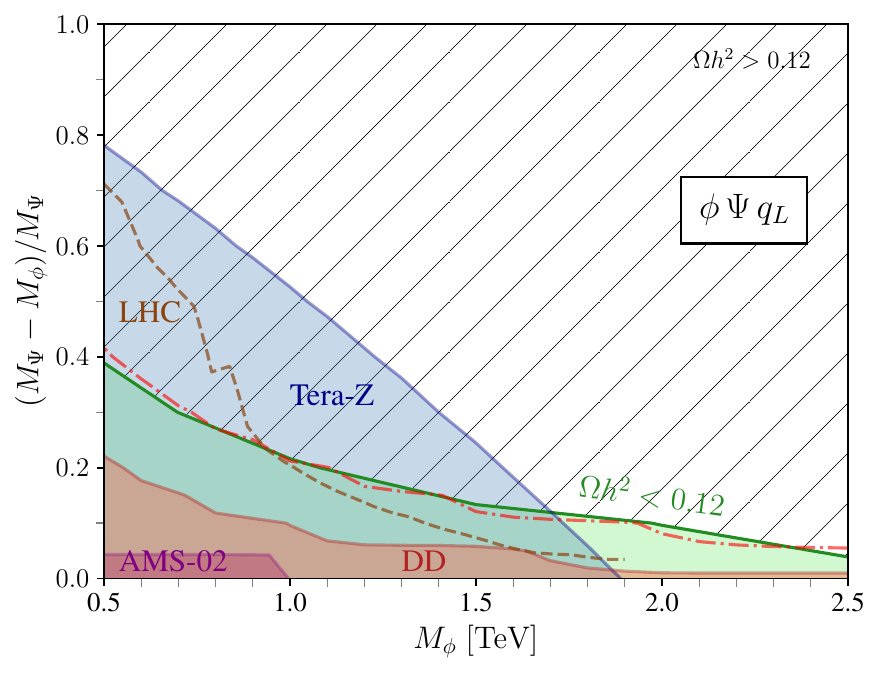}
    \caption{\emph{$\phi \Psi q_L$ case:} Parameter space with fixed mediator mass $M_\Psi = 2.5$ TeV (upper panel) and fixed portal coupling $y_Q=1.5$ (lower panel).
    Labels and regions same as in Fig.~\ref{fig:Fs_Sql}. Additionally, constraints from direct searches at LHC are shown by the dashed brown line in the lower panel~\cite{Arina:2025zpi}.}
    \label{fig:Ss_FqL}
\end{figure}

Moreover, colored fermionic mediators remain tightly constrained by LHC direct searches. Reinterpretations of existing analyses~\cite{CMS:2019zmd,ATLAS:2020xgt,ATLAS:2020syg,ATLAS:2021kxv,CMS:2021far} in Ref.~\cite{Arina:2025zpi} exclude fermionic mediators to third-generation quarks with masses up to $M_\Phi = 1.5$ TeV for a DM mass below $0.7$ TeV. Even stronger limits apply to mediators coupling to light quarks, reaching up to $M_\Phi = 1.9$ TeV for a DM mass below $0.8$ TeV for the second generation quarks. These constraints are illustrated by the brown dashed line in Fig.~\ref{fig:Ss_FqL} (lower panel), shown for the representative case of $\phi\Psi q_L^2$ portal. Looking ahead, HL-LHC projections are expected to strengthen these limits by roughly 30\%, enabling full coverage of the parameter space where the observed relic abundance is set via DM interactions with second-generation quark doublets~\cite{CidVidal:2018eel}. 
On the other hand, for the case of fermionic mediators to SM leptons, Ref. \cite{Guedes:2021oqx} sets the limit for a vector-like lepton (VLL) singlet decaying exclusively to a lepton and a vector boson at $M_{\Psi}=895$ GeV from direct LHC searches. Using the same limit from Ref.~\cite{Guedes:2021oqx}, and just rescaling the production cross-section for the VLL doublet, we obtain $M_{\Psi}=1500$ GeV. The projected HL-LHC exclusion limit for the singlet is $M_{\Psi}=1450$ GeV, with a similar improvement expected for the doublet as well.

In the following, we focus on the $t$-channel portals to $\psi=\{e_R,q_L,\ell_L\}$, omitting the cases with $\psi=\{u_R,d_R\}$. The portal with $\psi=u_R$ is only enhanced by the running effect in Eq.~\eqref{eq:2loop-FSu} if one assumes couplings to third generation, a scenario that is tightly constrained as mentioned above.
In the same fashion, we find again that the portal with $\psi=d_R$ yields a bottom Yukawa-suppressed running effect, and remains largely unconstrained at a Tera-$Z$ run.
A crucial distinction from the case of Majorana DM is that a scalar DM candidate has an unavoidable Higgs portal through the quartic coupling. This operator not only contributes to DM annihilation but also introduces a range of collider and (in)direct detection constraints, extensively studied in the literature~\cite{Yaguna_2009,Gonderinger_2010,Djouadi_2012,Cline_2013,kaplinghat2014direct,Duerr:2015aka,han2015new,gelmini2020scalar,vermeulen2021direct}. These typically impose stringent bounds, $\kappa \sim \mathcal{O}(0.01\, \text{--}\, 0.1)$ for $M_\phi \sim \mathcal{O}(1)$ TeV. For this reason, we focus on the limit of vanishing $\kappa$ in the analysis below.
\subsubsection{\texorpdfstring{$\phi\Psi e_R$}{phi-Psi-eR}}
\label{subsec:phi_Psi_eR}

\begin{figure}[t]
    \centering
    \vspace{-0.22cm}\includegraphics[width=\linewidth]{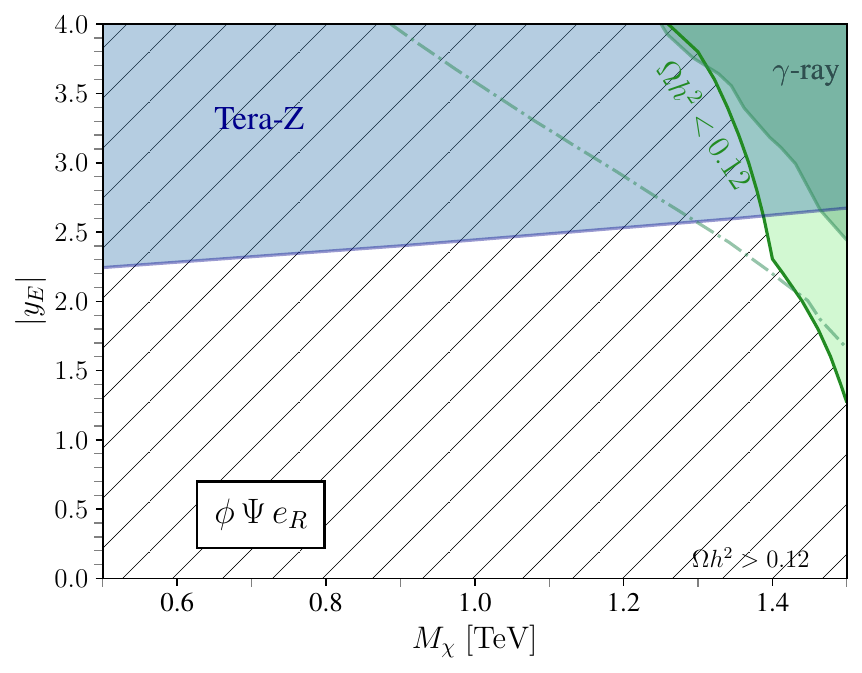}\vspace{-0.2cm}
    \includegraphics[width=\linewidth]{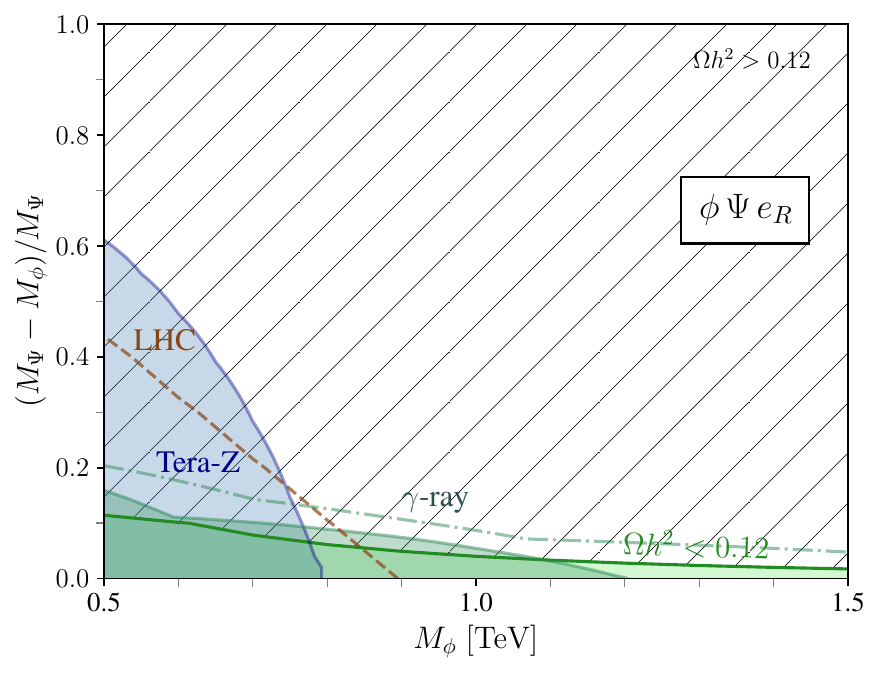}
    \vspace{-0.7cm}\caption{\emph{$\phi \Psi e_R$ case:} Parameter space with fixed mediator mass $M_\Psi = 1.5$ TeV (upper panel) and fixed portal coupling $y_E=2$ (lower panel). Labels and regions same as in Fig.~\ref{fig:Fs_Sql}. Additionally, the dark green region is excluded by $\gamma$-ray line searches at H.E.S.S.~\cite{HESS:2018cbt} and the dashed region corresponds to the projected sensitivity of CTAO~\cite{CTAO:2024wvb}. Constraints from direct searches at LHC are shown by the dashed brown line~\cite{Guedes:2021oqx}.
    }
    \label{fig:Ss_FeR}
\end{figure}

\begin{figure}[t]
    \centering
    \vspace{-0.15cm}\includegraphics[width=\linewidth]{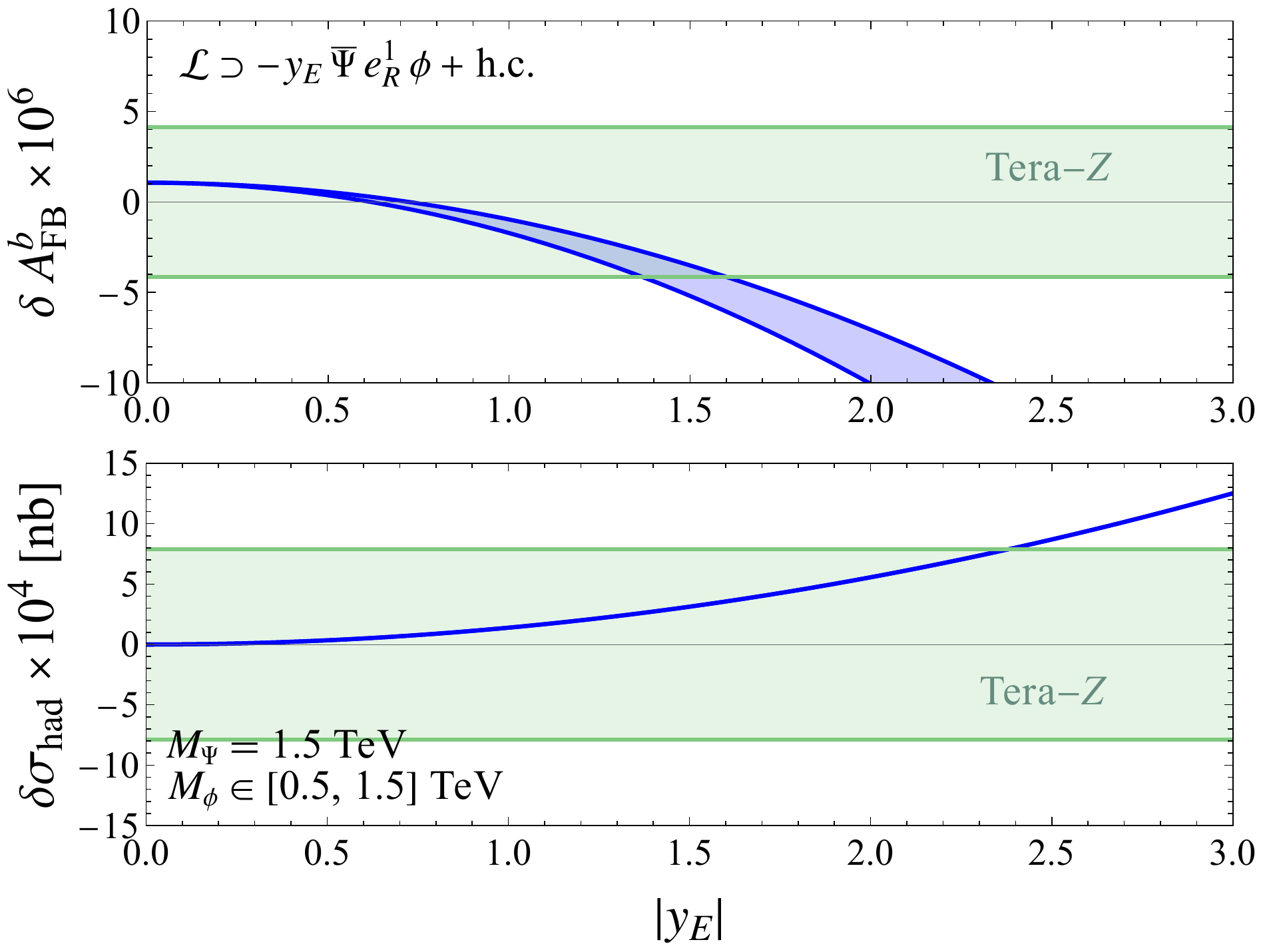}
    \vspace{-0.5cm}\caption{\emph{$\phi \Psi e_R$ case:} The variation of the most influential electroweak observables, $A_{\rm FB}^b$ and $\sigma_{\rm had}$, as a function of the portal coupling. 
    The green bands denote the Tera-$Z$ experimental sensitivity at $1\sigma$, while the blue bands reflect the variation of the dark matter mass $M_\phi \in [0.5, 1.5]$ TeV. The mediator mass is fixed to $M_\Psi = 1.5$ TeV, while $\kappa=0$.}
    \label{fig:SFe_EWPOs}
\end{figure}

The strongest constraints on the scalar DM coupled to the right-handed leptons originate from the interactions with the first generation 
\begin{equation}
      -\mathcal{L}\supset \kappa |\phi|^2 |H|^2 + \left(y_E \overline{\Psi}e_R^1 \phi +\mathrm{h.c.}\right)\,.
\end{equation}
The results shown in Figs.~\ref{fig:Ss_FeR} and~\ref{fig:SFe_EWPOs}, obtained with this flavour hypothesis, are quantitatively similar to the MFV one.  
The observables that drive Tera-$Z$ sensitivity in this case are $A_{\rm FB}^b$ and $\sigma_{\rm had}$, with the NP contribution arising from
\begin{equation}
    [C_{He}]_{11}\! \simeq\! \frac{g_1^2 |y_E|^2}{192\pi^2 M_\phi^2} 
    \!\left[\!\frac{x(29-7x)-16}{6(x-1)^3}-\frac{(3x-2)\log(x)}{(x-1)^4}\!\right]\!,
    \label{eq:CHe_SFe}
\end{equation}
where now $x=M_{\Psi}^2/M_{\phi}^2$. This contribution is always negative, leading to a correlated decrease in $A_{\rm FB}^b$ and increase in $\sigma_{\rm had}$, thereby giving a distinct signature at future $e^+e^-$ colliders. Moreover, the size of $C_{He}$ in Eq.~\eqref{eq:CHe_SFe} is at least a factor of three larger than in the Majorana DM case in Eq.~\eqref{eq:CHe_FSe}, implying that the right-handed lepton portal is substantially more constrained in the scalar DM scenario. Finally, complementary constraints on the target region can be set by current $\gamma$-ray line searches at H.E.S.S.~\cite{HESS:2018cbt} and by the projected sensitivity of CTAO~\cite{CTAO:2024wvb}.

\vspace{-0.3cm}\subsubsection{\texorpdfstring{$\phi\Psi q_L$}{phi-Psi-qL}}
\label{subsec:SFqL}

The Lagrangian for the portal to quark doublets reads
\begin{equation}
      -\mathcal{L}\supset \kappa |\phi|^2 |H|^2 +
      \left([y_Q]_i\overline{\Psi}\,q_L^i\phi +\mathrm{h.c.}\right)\,,
      \label{eq:Lag_SFq}
\end{equation}
where $i=\{1,2,3\}$ is a flavour index. As mentioned above, coloured mediators are significantly constrained by DD for scalar DM. In this model, couplings to first generation induce a tree-level $s$-channel scattering with $u,d$ quarks, and the third generation induces a Higgs penguin diagram enhanced by a factor $(M_\Psi/m_t)^2$ with respect to the Majorana DM case, where $m_t$ is the top-quark mass. For that reason, we focus on the scenario in which a Tera-$Z$ factory could be complementary with DD, \emph{i.e.}, assuming a coupling only to the second generation, $[y_Q]_i\propto \delta_{2i}$ in Eq.~\eqref{eq:Lag_SFq}.

\begin{figure}[t]
    \centering
    \includegraphics[width=\linewidth]{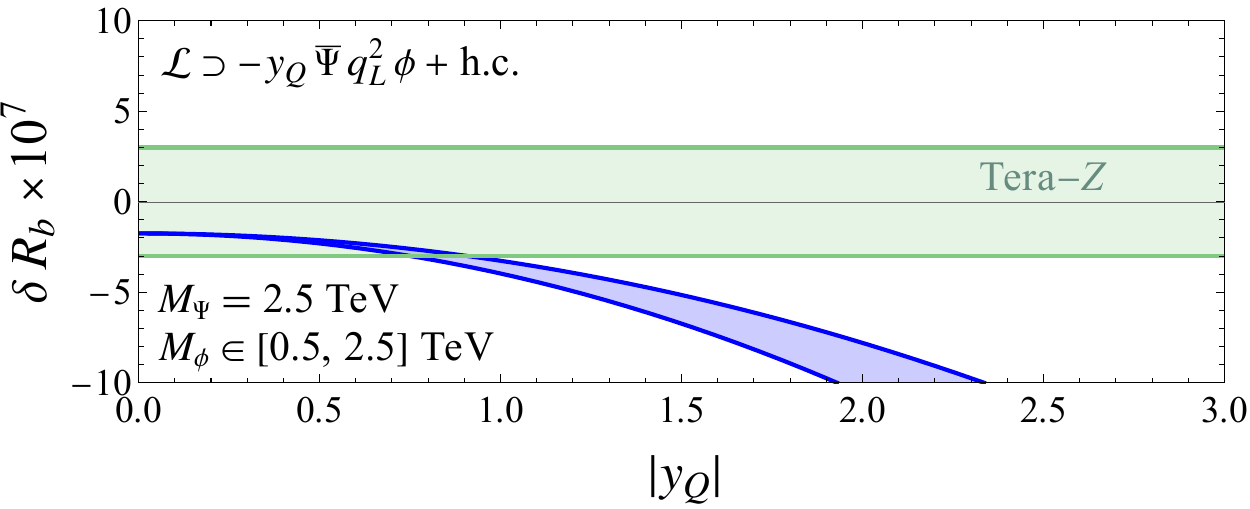}
    \caption{\emph{$\phi \Psi q_L$ case:} The variation of the most influential electroweak observable, $R_b$, as a function of the portal coupling. 
    The green bands denote the Tera-$Z$ experimental sensitivity at $1\sigma$, while the blue bands reflect the variation of the dark matter mass $M_\phi \in [0.5, 2.5]$ TeV. The mediator mass is fixed to $M_\Psi = 2.5$ TeV, while $\kappa=0$. NP contributions to other observables stay within their $1\sigma$ experimental uncertainties.}
    \label{fig:SFq_EWPOs}
\end{figure}

The observable with the highest pull in this model is $R_b$, as shown in Fig. (\ref{fig:SFq_EWPOs}). The main effects come from the coefficient $[C_{Hq}^{(3)}]_{22}$, which, with our flavor assumption reads
\begin{align}
    [C_{Hq}^{(3)}]_{22}&=-\frac{g_2^4}{160 \pi^2 M_\Psi^2}
    +\frac{g_2^2 |y_Q|^2}{2304\pi^2 M_\phi^2}\left[\frac{16+x(7x-29)}{(x-1)^3}\right. \nonumber\\
    &\left.+ \frac{6(2-3x)\log(x)}{(x-1)^4}\right],
\end{align}
with $[C_{Hq}^{(3)}]_{11,33}$ given by the same expression in the limit $y_Q=0$.

\subsubsection{\texorpdfstring{$\phi\Psi \ell_L$}{phi-Psi-ellL}}
\label{subsec:SFellL}

Finally, we introduce a vector-like lepton mediator $\Psi\sim(\mathbf{1},\mathbf{2},-1/2)$ with three degenerate generations:
\begin{equation}
    -\mathcal{L}\supset [y_L]_{ij}\overline{\Psi}^i\ell^j \phi + \mathrm{h.c.},
\end{equation}
imposing again a $U(3)_\ell$ symmetry on the coupling $y_L$, for comparison with model $\chi\Phi\ell_L$.

\begin{figure}[t]
    \centering
    \vspace{-0.15cm}\includegraphics[width=\linewidth]{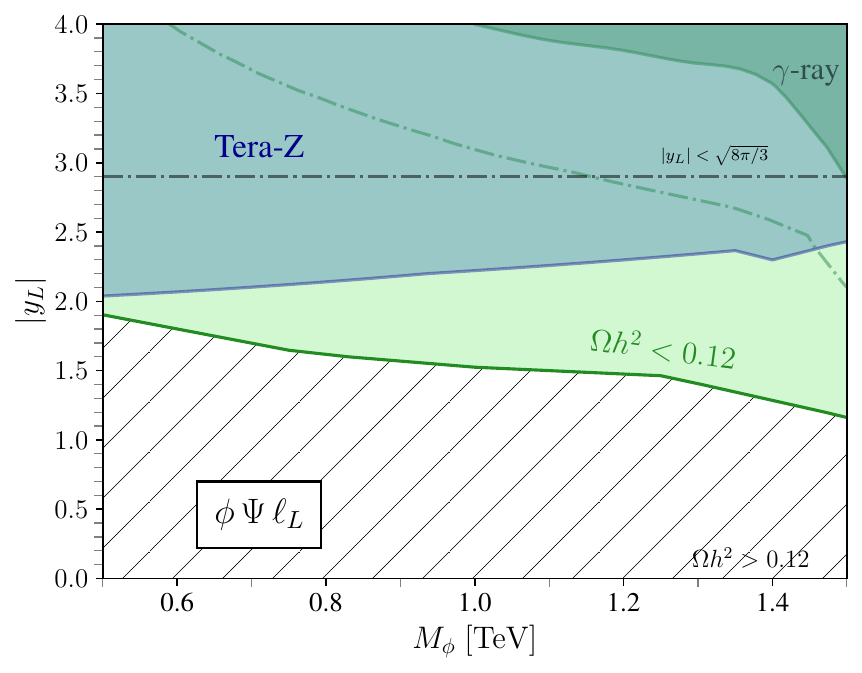}
    \includegraphics[width=\linewidth]{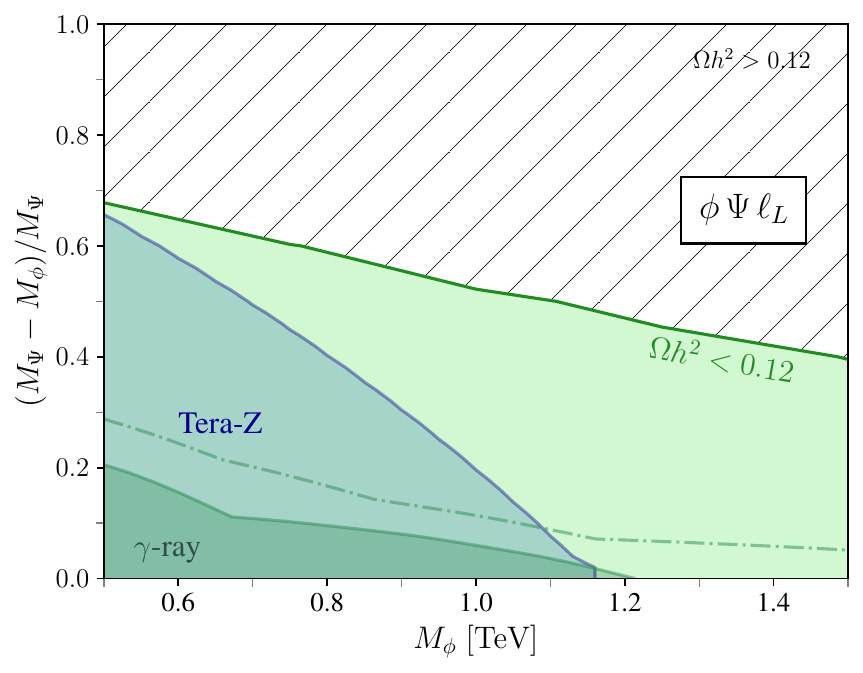}
    \caption{\emph{$\phi \Psi \ell_L$ case:} Parameter space with fixed mediator mass $M_\Psi = 1.5$ TeV (upper panel) and fixed portal coupling $y_L=2$ (lower panel).
    Labels and regions same as in Fig.~\ref{fig:Fs_Sql}.
    }
    \label{fig:Ss_FlL}
\end{figure}

The results in Fig.~\ref{fig:Ss_FlL} show how a Tera-$Z$ factory would constrain the target parameter space in a complementary way to current and future $\gamma$-ray line searches. For reference, we also show the limit on the coupling set by partial wave perturbative unitarity (see App.~\ref{sec:th_constraints}).

The main effects in this portal come again from the coefficients:
\begin{align}
    &\frac{1}{g_2^2}[C_{H\ell}^{(3)}]_{ii}=-\frac{1}{g_1^2}[C_{H\ell}^{(1)}]_{ii}=\frac{1}{2 g_2^2}[C_{\ell\ell}]_{1221}\simeq\\
    &\simeq\frac{|y_L|_i^2}{2304\pi^2 M_\phi^2}\left[\frac{16 + x ( 7 x -29)}{(x-1)^3}+\frac{6(2-3x)\log(x)}{(x-1)^4}\right]\nonumber,
\end{align}
which mainly affect $\Gamma_Z$ and $A_{\mathrm{FB}}^b$ in a correlated decreasing deviation from the SM prediction. Note that, in contrast to the $\chi\Phi\ell_L$ scenario, we do not find the aforementioned cancellation in MFV because there is no $|y_L|^4$ contribution to $C_{\ell\ell}$ in this case.

\begin{figure}[t]
    \centering
    \includegraphics[width=\linewidth]{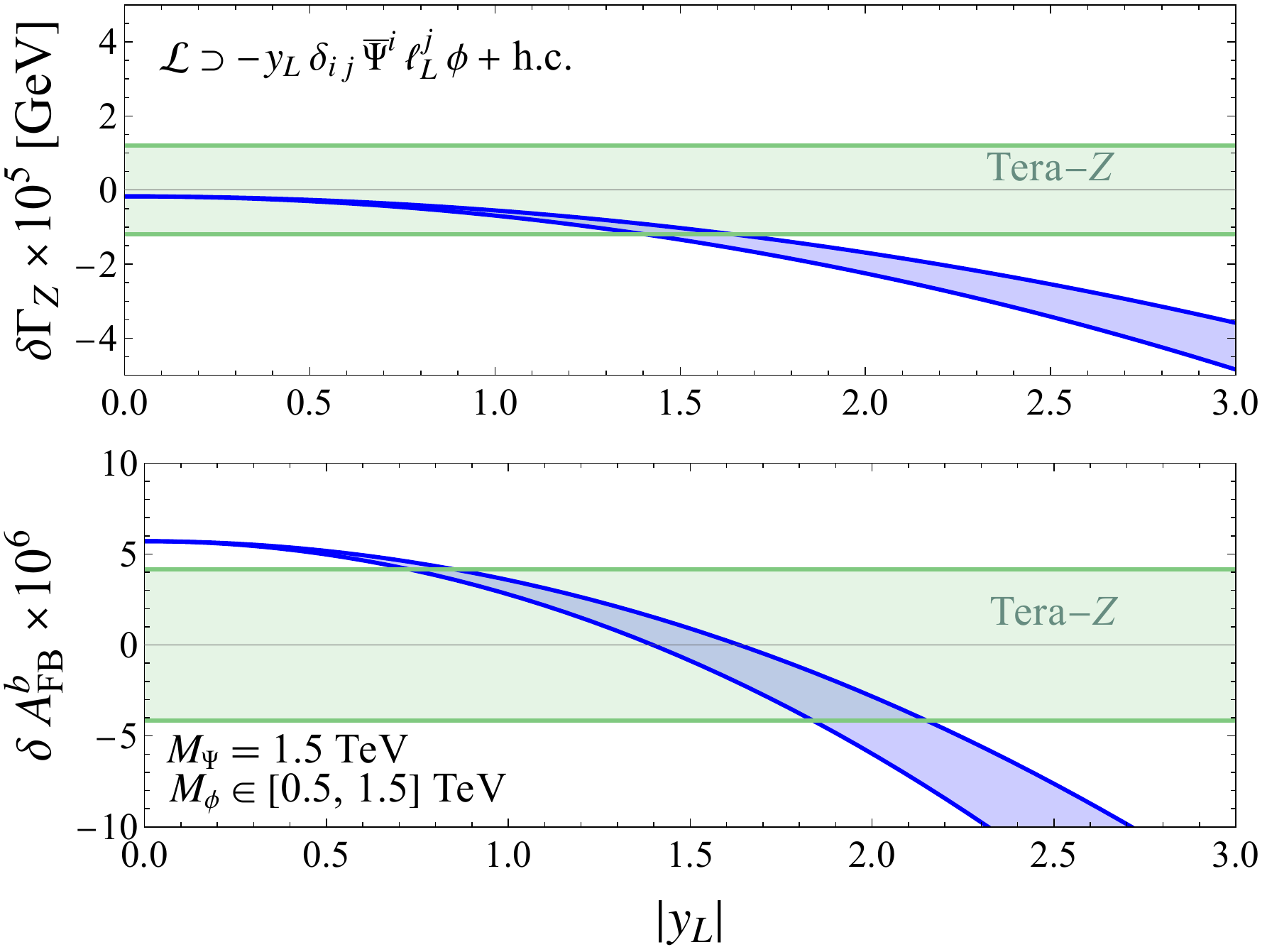}
    \caption{\emph{$\phi \Psi \ell_L$ case (MFV):} The variation of the most influential electroweak observables, $\Gamma_Z$ and $A_{\rm FB}^b$, as a function of the portal coupling. 
    The green bands denote the Tera-$Z$ experimental sensitivity at $1\sigma$, while the blue bands reflect the variation of the dark matter mass $M_\phi \in [0.5, 1.5]$ TeV. The mediator mass is fixed to $M_\Psi = 1.5$ TeV, while $\kappa=0$.}
    \label{fig:SFell_EWPOs}
\end{figure}

\section{Conclusions} 
\label{sec:conclusions}

The next-generation circular $e^+e^-$ collider, such as FCC-ee and CEPC, will enable electroweak precision measurements with unprecedented accuracy, offering powerful sensitivity to new physics through small deviations from Standard Model predictions. 

In this work, we have analysed a class of dark matter models where the DM particle is a Standard Model singlet and requires 
a mediator to couple to the visible sector, so-called $t$-channel portals. Since the DM as well as the mediator are odd under a $\mathbb{Z}_2$ symmetry which ensures the dark matter stability, all effects in the visible sector are induced only at one-loop order. Despite the minimal setup, this framework captures a broad class of dark sectors, allowing for a systematic investigation of their imprint on electroweak precision observables at a future Tera-$Z$ run.

Our results show that such a run can indirectly probe the presence of DM and yield valuable insight into its underlying properties. In particular, we demonstrated that a Tera-$Z$ program could achieve sensitivity to such scenarios, in some cases unexpectedly competitive with dedicated dark matter experiments like DARWIN and CTAO. Moreover, in leptonic portals, electroweak precision measurements often provide the only available probe.

We also identified which electroweak precision observables would benefit the most from improved Standard Model predictions, offering a clear target for future theory efforts aimed at enhancing the discovery potential of a Tera-$Z$ run in the context of DM searches. Remarkably, we find that even certain two-loop contributions may become accessible thanks to the precision reach, motivating a dedicated two-loop matching and running program in the EFT community. In several cases, such higher-order effects can dominate the signal and carry key discriminatory power among different portal realizations.

This article represents the first study of the sensitivity of the Tera-$Z$ program to an important class of models where the leading effects arise at the one-loop level. Importantly, we found that different portals give rise to distinct patterns of correlations, making a Tera-$Z$ an essential tool for disentangling the nature of the DM and the corresponding mediator.

The framework presented here can be extended in several directions, including alternative flavor structures, additional degrees of freedom, and more complete UV scenarios. Nevertheless, already in its minimal form, it underscores the power of future electroweak precision measurements to shed light on dark sectors beyond the current experimental horizon. 

\section*{Acknowledgments} 
\label{sec:acknowledgment}

We are grateful to Mathias Becker for useful discussions. This work received funding by the INFN Iniziative Specifiche AMPLITUDES and APINE and from the European Union’s Horizon 2020 research and innovation programme under the Marie Sklodowska-Curie grant agreements n. 860881 – HIDDeN, n. 101086085 – ASYMMETRY. This work was also partially supported by the Italian MUR Departments of Excellence grant 2023-
2027 “Quantum Frontiers”, the European Union – Next Generation EU, and the Italian Ministry of University and Research (MUR) via the PRIN 2022 project n. 2022K4B58X – AxionOrigins.

\newpage
\appendix
\section*{Appendix}
\renewcommand{\thesubsection}{\thesection\Alph{subsection}}

\subsection{Theoretical constraints} 
\label{sec:th_constraints}

In addition to experimental constraints—such as those arising from electroweak precision observables at Tera-$Z$ and direct detection experiments—we can also impose theoretical bounds on the viability of the models discussed above.

A relevant example is the scale at which the renormalization group evolution of the new physics couplings leads to a Landau pole. To illustrate this, we consider the models $\phi\Psi q_L$ and $\chi\Phi e_R$. For the $\phi \Psi q_L$ scenario, the beta function of the $y_Q$ coupling (see Eq.~\eqref{eq:Lag_SFq}) is approximately given by
\begin{equation}
\beta_{y_Q} \simeq -\frac{7 g_3^2 y_Q}{12\pi^2} + \frac{33 y_Q^3}{64\pi^2}\,,
\end{equation}
where we neglect electroweak gauge couplings ($g_1$, $g_2$) and assume interactions to only one generation.

\begin{figure}[t]
    \centering
    \includegraphics[width=\linewidth]{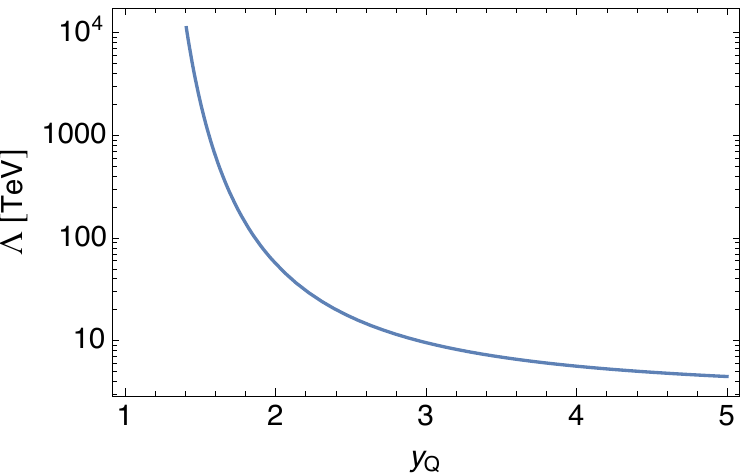}
    \caption{Scale $\Lambda$ at which a Landau pole appears, as a function of the coupling $y_Q$ defined at the reference scale $\mu = 3$ TeV, in the $\phi \Psi q_L$ model.}
    \label{fig:landau_Ss_Fq}\vspace{0.3cm}
    \includegraphics[width=\linewidth]{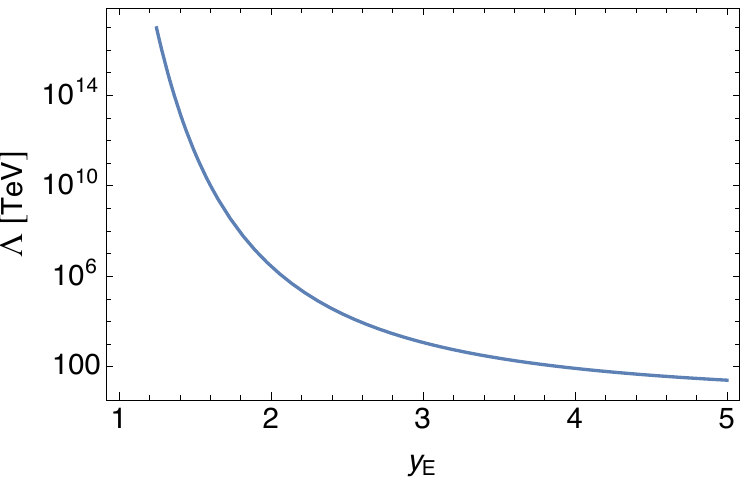}
    \caption{Scale $\Lambda$ at which a Landau pole appears, as a function of the coupling $y_E$ defined at the reference scale $\mu = 3$ TeV, in the $\chi\Phi e_R$ model.}
    \label{fig:landau_Fs_Ss}
\end{figure}

In Fig.~\ref{fig:landau_Ss_Fq}, we show the scale at which a Landau pole develops as a function of the initial value of $y_Q$ defined at the mediator scale, $M_\Psi = 3$ TeV. Requiring that the theory remains perturbative up to at least $10^3$ TeV sets an upper bound on $y_Q$, with important phenomenological consequences. In particular, for $\kappa = 0$, the $\phi\Psi q_L$ model can only reproduce the observed dark matter relic abundance in the limit $M_\phi\simeq M_\Psi$ without violating this perturbativity condition.

For comparison, we also show the results for the $\chi\Phi e_R$ model. The beta function for the $y_E$ coupling reads:
\begin{equation}
     \beta_{y_E}= -\frac{g_1^2 y_E}{4\pi^2}+\frac{3 y_E^3}{32\pi^2}\,.
\end{equation}
The results for the Landau pole are shown in Fig.~\ref{fig:landau_Fs_Ss}.

In addition, further theoretical constraints can be derived from perturbative unitarity~\cite{Itzykson:1980rh,Chanowitz:1978mv,DiLuzio:2017chi}. The resulting bounds are summarized in Tab.~\ref{tab:unitarity}.

\begin{table}[h]
\renewcommand{\arraystretch}{2} 
\setlength{\tabcolsep}{12pt} 
    \centering
   \begin{tabular}{|c|c|}
  \hline
  \textbf{Coupling} & \textbf{ Unitarity bound} \\
  \hline
  $y_L$ & $\sqrt{8\pi/3}$ \\
  \hline
  $y_Q$ & $\sqrt{8\pi/9}$ \\
  \hline
  $y_E$ & $\sqrt{16\pi/3}$ \\
  \hline
  $y_U$ & $\sqrt{16\pi/9}$ \\
  \hline
\end{tabular}
    \caption{Naive bounds set by partial wave perturbative unitary, assuming flavor universality.} 
    \label{tab:unitarity}
\end{table}

\subsection{Direct Detection}
\label{app:DD}
Considering a Majorana singlet DM candidate, the effective Lagrangian of interaction with quarks can be written as \cite{Hisano:2010ct,Hisano:2015bma,Hisano:2018bpz,Gondolo:2013wwa,Arcadi:2023imv}:
\begin{align}
    \mathcal{L}&= \sum_{q=u,d,s}c_A^q \,\overline{\chi}\gamma_\mu\gamma_5\chi \overline{q}\gamma^\mu\gamma_5 q  + \sum_{q=u,d,s} c_S^q  m_q \overline{\chi}\chi \overline{q}q \nonumber\\
    &+ \sum_{q=c,b,t} c_G^{(q)} \overline{\chi}\chi G^a_{\mu\nu}G^{a\mu\nu}\nonumber\\
    &+\sum_{q=u,d,s} \big(g_q^{(1)}\frac{\overline{\chi}i\partial^\mu\gamma^\nu\chi}{m_\chi}
    +g_q^{(2)}\frac{\overline{\chi}i\partial^\mu i\partial^\nu\chi}{m_\chi}
    \big)\mathcal{O}_{q\mu\nu}^{(2)}\nonumber\\
    &+\sum_{q=c,b,t} \big(g_G^{(1,q)}\frac{\overline{\chi}i\partial^\mu\gamma^\nu\chi}{m_\chi}
    +g_G^{(2,q)}\frac{\overline{\chi}i\partial^\mu i\partial^\nu\chi}{m_\chi}
    \big)\mathcal{O}_{G\mu\nu}^{(2)}\,.
    \label{eq:DD_majorana_lag}
\end{align}
The twist-2 operators $\mathcal{O}_{q\mu\nu}^{(2)},\,\mathcal{O}_{G\mu\nu}^{(2)}$ with quarks and gluons can be found in Ref.~\cite{Hisano:2015bma}.
In the non-relativistic limit, this Lagrangian induces the following effective DM-nucleon coupling $f_N$ \cite{Hisano:2015bma,Hisano:2010ct,Gondolo:2013wwa}:
\begin{align}
    f_N/m_N&=\frac{2}{27}\sum_{q=u,d,s} f_{Tq}^{(N)} c_S^{(q)} -\frac{8\pi}{9\alpha_s}f_{TG}\sum_{q=c,b,t} (c_G^{(q)} - \frac{ \alpha_s}{12\pi}c_S^{(q)} ) \nonumber\\
    &+ \frac{3}{4}G(2)\sum_{q=c,b,t}(g_G^{(1,q)}+g_G^{(2,q)})\nonumber\\
    &+ \frac{3}{4}\sum_{q=u,d,s} (g_q^{(1)}+g_q^{(2)})(q(2)+\overline{q}(2))\,,
\end{align}
where $m_N$ is the mass of the nucleon and $f_{Tq}, f_{TG}, G(2)$ and $q(2)+\overline{q}(2)$ are nucleon form factors that can be found in 
\cite{Hisano:2015bma,Mohan:2019zrk,Hill:2014yxa,Young:2009zb,PhysRevD.87.034509,Owens:2012bv}.
We compute the spin-independent cross-section of DM with a nucleon as:
\begin{equation}
    \sigma_{\mathrm{SI}}^{(N)}=\frac{4\,\mu_N^2}{\pi}f_N^2,
\end{equation}
with $\mu_N=\frac{m_\chi m_N}{m_\chi+m_N}$ the reduced mass of the DM-nucleon system. Likewise, we compute the spin-dependent cross-section as  \cite{Mohan:2019zrk,Arcadi:2023imv}:
\begin{equation}
    \sigma_{\mathrm{SD}}^{(N)}=\frac{12\,\mu_N^2}{\pi}\Big[\sum_{q=u,d,s} c_A^{(q)}\Delta q^N)\Big]^2\,.
\end{equation}
In the case of a real scalar DM candidate, we can write the effective Lagrangian as \cite{Hisano:2015bma,Arcadi:2023imv}:
\begin{align}
    \mathcal{L}&= \sum_{q=u,d,s} c_S^q m_q \phi^2 \overline{q}q
    + \sum_{q=c,b,t} c_G^{(q)} \phi^2 G_{\mu\nu}^a G^{a\mu\nu}\nonumber\\
    &+\sum_{q=u,d,s} 
    \frac{g_q}{m_\phi^2}
    \phi i\partial^\mu i\partial^\nu\phi\;\mathcal{O}_{q\mu\nu}^{(2)}\nonumber\\
    &+\sum_{q=c,b,t} 
    \frac{g_G^{(q)}}{m_\phi^2}
    \phi i\partial^\mu i\partial^\nu\phi\;\mathcal{O}_{G\mu\nu}^{(2)}\,.
    \label{eq:DM_EFT_Lag}
\end{align}
This Lagrangian, in turn, induces the following DM-nucleon coupling \cite{Hisano:2010ct,Gondolo:2013wwa,Arcadi:2023imv}:
\begin{align}
    f_N/m_N&=\frac{2}{27}\sum_{q=u,d,s} f_{Tq}^{(N)} c_S^{(q)} -\frac{8\pi}{9\alpha_s}f_{TG}\sum_{q=c,b,t} (c_G^{(q)} - \frac{ \alpha_s}{12\pi}c_S^{(q)} ) \nonumber\\
    &+ \frac{3}{4}G(2)\sum_{q=c,b,t}g_G^{(q)}\nonumber\\
    &+ \frac{3}{4}\sum_{q=u,d,s} g_q(q(2)+\overline{q}(2))\,.
\end{align}
Finally, we write the spin-independent cross-section of DM with a nucleon as:
\begin{equation}
    \sigma_{\mathrm{SI}}^{(N)}=\frac{1}{\pi}\Big(\frac{m_N}{m_N+m_\phi}\Big)^2f_N^2
    \,.
\end{equation}


\bibliographystyle{JHEP}
\bibliography{references.bib}

\end{document}